\pdfoutput=1

\documentclass[11pt]{article}

\usepackage[final]{acl}

\usepackage{times}
\usepackage{latexsym}
\usepackage{subfigure}

\usepackage[T1]{fontenc}

\usepackage[utf8]{inputenc}

\usepackage{microtype}

\usepackage{inconsolata}

\usepackage{graphicx}

\usepackage{amsmath}
\usepackage{amssymb}
\usepackage{algorithm}
\usepackage{algorithmic}
\usepackage{bbm}
\usepackage{forest}
\usepackage{makecell}
\usepackage{multirow}
\usepackage{longtable}

\title{Automatic Evaluation Metrics for Artificially Generated Scientific Research}

\author{
 \textbf{Niklas Hoepner\textsuperscript{1}},
 \textbf{Leon Eshuijs\textsuperscript{2}},
 \textbf{Dimitrios Alivanistos\textsuperscript{2}},
 \textbf{Giacomo Zamprogno\textsuperscript{2}},
\\
 \textbf{Ilaria Tiddi\textsuperscript{2}}
\\
 \textsuperscript{1} University of Amsterdam,
 \textsuperscript{2} Vrije Universiteit
\\
 \small{
   \textbf{Correspondence:} \href{mailto:email@domain}{n.r.hopner@uva.nl}
  }
}

\begin{document}
\maketitle
\begin{abstract}

Foundation models are increasingly used in scientific research, but evaluating AI-generated scientific work remains challenging. Expert reviews are costly, while large language models (LLMs) as proxy reviewers have proven to be unreliable. To address this, we investigate two automatic evaluation metrics, specifically citation count prediction and review score prediction. We parse all papers of OpenReview and augment each submission with its citation count, reference, and research hypothesis. Our findings reveal that citation count prediction is more viable than review score prediction, and predicting scores is more difficult purely from the research hypothesis than from the full paper. Furthermore, we show that a simple prediction model based solely on title and abstract outperforms LLM-based reviewers, though it still falls short of human-level consistency\footnote{\url{https://github.com/NikeHop/automatic_scientific_quality_metrics}}\footnote{\url{https://github.com/NikeHop/OpenReviewParser}}.
\end{abstract}

\section{Introduction}

Advances in foundation models \cite{palm_e,gpt4} have increased interest in their potential to enhance various stages of the research process \cite{llm_chemical_research,coauthor}, including research hypothesis generation \cite{scimon,research_agent,NLP100}, writing assistance \cite{writing_assistant1,writing_assistant2}, and peer review \cite{reviewer_gpt,review_generation1}. Their application to literature-based generation of scientific content has been particularly notable, producing research hypotheses that domain experts consider more novel than those generated by humans \cite{NLP100}. While much of the focus has been on generating research hypotheses \cite{NLP100,research_agent,scimon}, fewer studies have explored the generation of complete scientific papers or multiple steps of the research process \cite{sakana,machine_learning_research_agent}.

A key challenge in this area is evaluating the quality of generated scientific content \cite{research_agent,sakana,NLP100}. Most studies use a mix of domain expert assessments and LLM-based reviews \cite{research_agent,scimon,hypothesis_generation}. However, expert evaluations are costly and time-consuming, limiting them to a subset of methodologies, often selected based on LLM reviewer results \cite{hypothesis_generation}. While early studies showed promise for LLMs in reviewing papers \cite{research_agent,sakana}, recent findings reveal their reliability issues, with some cases showing LLM reviewers performing no better than random guessing compared to human evaluations of novel research hypotheses \cite{NLP100}.

An automatic evaluation metric should be efficient to compute, reliably indicate scientific quality and generalise to unseen future work. The field of scholarly document quality prediction (SDQP) \cite{multischubert} has studied the problem of predicting review scores and citation counts \cite{peer_read,cimate} as proxy metrics. Domain experts reviews are the gold standard for evaluating the scientific quality of a paper and citation counts measure a paper's influence on the scientific community. However, limited data availability has constrained review score prediction \cite{review_dataset_overview}, while citation prediction typically relies on factors beyond paper content such as paper metadata \cite{citation_prediction_sciencometrics}. Prior work has not studied the challenge of predicting scientific quality based on research hypotheses alone and assessing generalisation to future work.

To study both prediction problems, we propose data models for scientific papers and reviews to parse all OpenReview submissions\footnote{\url{https:/openreview.net}} into a unified format, further annotating submissions with citation counts and research hypotheses. This allows us to analyse the relationship between review and citation scores, evaluate their predictability, and compare them with LLM-based review systems.

In summary, our contributions are:

\begin{itemize}
\itemsep0em 
    \item We propose citation count and review score prediction as automatic evaluation metrics for AI-generated scientific content.
    \item We parse all OpenReview submissions into a unifying format, augmenting them with additional metadata.
    \item Demonstrate that a simple score prediction model is more consistent with human reviews than LLM-based reviewers. 
\end{itemize}

\section{Related Work}

\subsection{AI generated Science}

Advancements in instructable generative models \cite{gpt4,stable_diffusion_3} have sparked interest in their use for scientific content generation \cite{scimon,sakana,NLP100,chemistry_tools,writing_assistant1}. Large Language Models (LLMs) have been explored for tasks such as  research hypothesis generation \cite{scimon,research_agent,NLP100}, paper drafting \cite{writing_assistant1,writing_assistant2}, and even experimental design \cite{chemistry_tools, machine_learning_research_agent}. Most studies focus on specific subproblems. Notable exceptions include \citet{sakana}, who apply LLMs to the entire research process, and \citet{machine_learning_research_agent}, who focus on machine learning research but exclude report writing.

A key challenge in this domain is evaluating generated outputs, such as the novelty or feasibility of a research hypothesis, which often requires time-consuming, costly reviews by domain experts \cite{NLP100}. To address this, some researchers have used LLM-based evaluations as substitutes for human judgment \cite{research_agent,hypothesis_generation,hypothesis_generation2}, with some even replacing human evaluators in certain cases entirely \cite{sakana}. However, recent studies question the reliability of LLMs as evaluators compared to human reviewers \cite{NLP100}. We propose citation count and review score prediction as alternative evaluation metrics and compare them to LLM-based systems.

\subsection{Automated Peer Reviewing}

With the increase in scientific paper submissions, interest in automating aspects of the peer review process has grown \cite{automated_review_process}, spanning tasks like reviewer assignment \cite{malicious_bidding} and review generation \cite{review_generation1,review_generation2,review_generation3}. An overview of the tasks and related datasets is provided in \citet{review_dataset_overview}.

While, automatically generated textual reviews  \cite{review_generation2} can provide valuable feedback to authors, they are unsuitable as standalone review metrics, as it is difficult to determine which method generates more valuable research papers. To develop automatic evaluation metrics, we focus on the task of \textit{Review Score Prediction} (RSP) \cite{peer_read}. Another related task, \textit{Paper Decision Prediction} (DSP) \cite{peer_assist,automated_review_process}, could also contribute to evaluation metrics but offers less granularity. While \citet{peer_read} explore review score prediction, their work is limited to a small dataset (PeerRead,<1000 submissions) and does not address temporal generalization. Follow up work focuses on improving prediction accuracy by dealing with the limited data by learning from unlabelled data \cite{review_score_prediction_unlabeled} or adding intermediate tasks for finetuning \cite{review_score_prediction_intermediate}.

As noted by \citet{review_dataset_overview}, most studies create their own datasets, many of which are derived from OpenReview. Alternative sources of reviews, such as F1000\footnote{\url{https://f1000research.com/}} and PeerJ\footnote{\url{https://peerj.com/}} cover a more diverse range of topics but lack an anonymous review process \cite{other_dataset}. The lack of a standardized, updated dataset with rich metadata limits method comparisons and progress in automating reviews. Our work addresses this by providing data models for scientific papers and reviews to unify all OpenReview submissions.

\subsection{Citation Count Prediction}

Citation count prediction approaches differ based on the information used \cite{citation_prediction_sciencometrics}. Most models combine citation history \cite{citation_history1}, metadata (e.g., authors, h-index) \cite{citation_metadata1}, and paper content \cite{schubert}. For newly generated content, without citation history or metadata, only content-based models are applicable to build evaluation metrics. Early studies used n-gram and term-frequency features from titles and abstracts, with small datasets (~3,000 papers) and short time spans \cite{early_work_citation_prediction1, early_work_citation_prediction2}. Recent efforts scaled dataset sizes (~30,000–40,000 papers) and incorporated embeddings from large language models to improve prediction performance \cite{schubert, embeddings_citation_count_prediction}. Some studies show significant benefits from using full papers, while others report only marginal improvements \cite{embeddings_citation_count_prediction}.

These conflicting findings may stem from differences in datasets and target metrics employed by each study. Such variations complicate direct comparisons across works, hindering the overall progress of the field. Furthermore, few approaches explicitly account for generalization over time. Notably, many studies use validation sets with publication dates that are only minimally separated from those in the training set, with a maximum gap of one month \cite{newly_published_citation_prediction, cimate}.

Few studies have explored the relationship between citation counts and peer-review scores. While prior work shows that reviews can improve citation prediction models \cite{citation_prediction_review_text1, citation_prediction_review_text2}, our focus is on their correlation. A strong correlation would suggest both metrics capture similar aspects of a paper's quality. \citet{findings_openreview} found a weak positive correlation between citation counts and review scores for ICLR submissions from 2017 to 2019, which we extend to the entire OpenReview dataset.

\section{Score Prediction}

\begin{figure}[t]
    \centering
    \includegraphics[width=\linewidth]{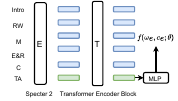}
    \caption{Architecture of the context model in case of the full paper representation. The paper representation is green (TA=title abstract) and the context representation is blue (RW=Related Work, M=Methodology, E\&R=Experiments and Results, C=Conclusion).}
    \label{fig:transformer_architecture}
\end{figure}

To address the challenge of evaluating the quality of scientific content generated by LLMs, we study the problem of predicting quality scores of scientific papers. As proxies for quality, we utilize citation counts and review scores. We consider two primary tasks: (1) pairwise ranking, where the objective is to predict which of two papers has a higher quality score, and (2) regression, where the goal is to predict the quality score of a single paper. The pairwise ranker can generate a ranking of generated content by comparing items in sequence, using either a round-robin format where each item is compared against every other item, or a Swiss tournament system where items are matched based on their current rankings \cite{NLP100}.

Formally, the dataset is represented as  $D = \{(\omega,c,s)\}^{N}_{i=1}$, where $\omega$ denotes the representation of a paper, $c$ refers to paper context such as references and $s$ is the associated quality score. In this work, we investigate representing papers via different part of the paper ranging from title and abstract to their research hypothesis. Given a scientific text $\omega$ or $c$, we compute embeddings using the SPECTER2 model\footnote{\url{https://huggingface.co/allenai/specter2_base}} with the regression adapter \cite{specter}. SPECTER2 is specifically designed to compute embeddings for tasks requiring high-quality representations of scientific text. When the input text exceeds the context length supported by the embedding model, we segment the text into sentences using the NLTK sentence tokenizer \cite{nltk}. For each chunk, we compute its embedding independently and subsequently average the embeddings to obtain a fixed-size representation of the entire text. We freeze the parameters of the embedding model and denote embedded scientific text via subscript $e$ ($\omega_{e}$,$c_{e}$).

Let $d_i = (\omega_{e,i}, c_{e,i}, s_i)$ denote the input data for a single paper. The score model $f_{\theta}$ is trained to predict the mapping from the target score by minimizing two objectives:

1. \textbf{Pairwise Comparison:} We minimize the binary cross-entropy loss between score differences:

\begin{equation}
L(d_1, d_2; \theta) = x \log y - (1 - x) \log(1 - y),
\end{equation}

where $ x = \mathbbm{1}_{s_1 > s_2}$ is a binary indicator for the score comparison, and $ y = \sigma(f_{\theta}(d_1) - f_{\theta}(d_2)) $ represents the predicted probability that paper $d_{1}$ has a higher score than paper $d_{2}$.

2. \textbf{Direct Score Prediction}: We minimize the L1 distance to learn the direct mapping of the input to the score:

\begin{equation}
L(\omega_e, c_e, s; \theta) = |f_{\theta}(\omega_e, c_e) - s|.
\end{equation}

For models without contextual information, \( f \) is implemented as an MLP with a single hidden layer. When context is included, it is represented as a sequence of embeddings (e.g., SPECTER2 embeddings derived from the title and abstract of the references). The context embeddings \( c_e \) are combined with the paper representation \( \omega_e \) to form a sequence of embeddings which is then processed through a one-layer Transformer encoder \cite{attention}. The resulting paper representation embedding is then passed through a MLP similarly to predict the target score (see Figure \ref{fig:transformer_architecture}). We look at two types of contexts. In the first, the paper is represented by its title and abstract, with the remaining sections serving as the contextual information. In the second approach, to incorporate references, the paper is still represented by its title and abstract, while the context is formed by the title and abstract of the references. We choose the following target scores for the each score type.

\textbf{Citation Count:}
For the citation count, we use the average citations per month as the prediction metric. The score is calculated by dividing the total citation count at the time of model training by the number of months since the paper was published. Given that the distribution of citation scores is highly skewed—where many papers have only a few citations while others receive a significantly larger number—we predict the logarithm of the average citations per month \cite{schubert,multischubert}. 

\textbf{Review Scores:}
For the available data on double-blind peer-reviews, each paper is associated with multiple reviews, which can vary in format depending on the conference the paper is submitted to. Reviews often include an overall decision, a review score, and evaluations of specific aspects of the paper, such as clarity and impact. We map all the review data to our review data model (Figure \ref{fig:review_schema} in Appendix \ref{app:data_model}). From the available scores, we predict the average overall review score and impact score.

More complex models are possible. However, we found larger models to overfit on the training set and not generalize to the more recent papers of the test set despite applying common regularization techniques \cite{dropout}. Further training details are described in Appendix \ref{app:training_details}.

\section{Datasets}

\begin{figure}
    \centering
    \includegraphics[width=\linewidth]{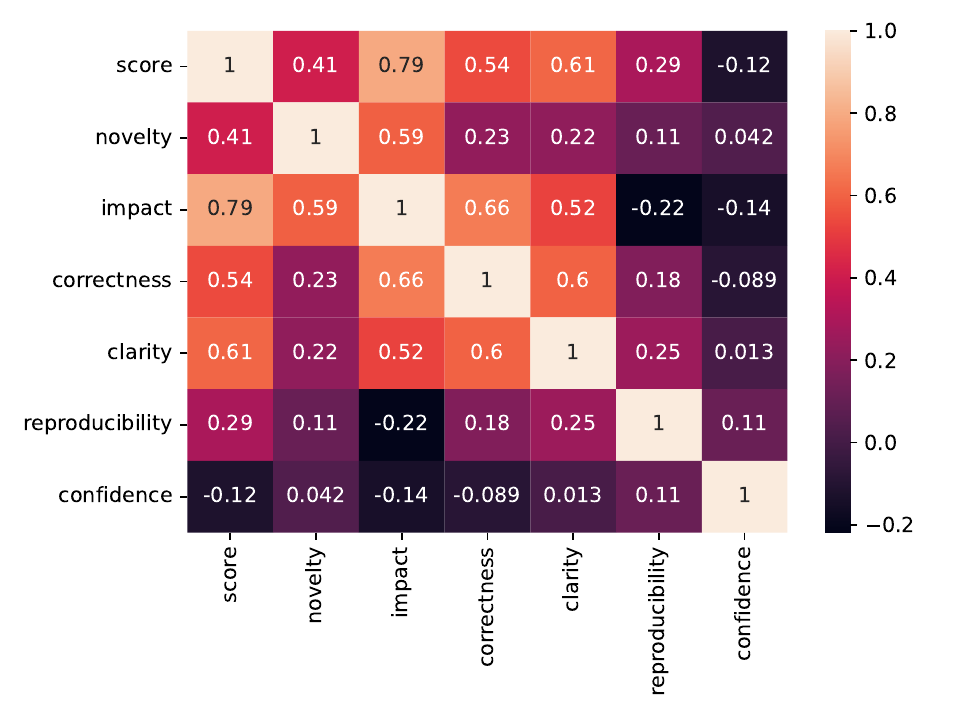}
    \caption{Pearson correlation heat map for the different dimensions of our unified review data model.}
    \label{fig:review_score_corr}
\end{figure}

Our work builds on the widely used ACL-OCL dataset \cite{acl_ocl}, enriched with updated citation counts, title and abstract of references and annotated research hypotheses. We created to the best of our knowledge largest dataset of OpenReview submissions that exists and augmented submissions with the same metadata as the ACL-OCL dataset \footnote{\url{https://huggingface.co/datasets/nhop/scientific-quality-score-prediction}}.
 
\textbf{OpenReview:} OpenReview\footnote{\url{https://openreview.net/}} serves as a valuable resource for linking scientific paper submissions with their corresponding peer-review assessments and has been a key source for datasets used in studies of the peer-review process \cite{review_dataset_overview}. However, a significant challenge lies in the lack of a unified format of reviews from different venues. For instance, while many workshops include only an overall score, ICLR-2023 includes five distinct fields, such as clarity and novelty (see Table \ref{tab:iclr_2023_review_fields} in Appendix \ref{app:openreview_dataset}). To address this, we developed a unified data model for reviews and manually mapped the fields from each venue to our standardized schema (see Table \ref{tab:review_field_harmonisation} in Appendix \ref{app:openreview_dataset}).

For each submission, we processed the PDF documents using GROBID \footnote{\url{https://github.com/kermitt2/grobid}}. To flexibly condition on different parts of an academic paper we train a section classifier that takes a paragraph of an academic text as input and classifies it as introduction, background, methodology, experiments \& results or conclusion. More details on the training of the section classifier can be found in Appendix \ref{app:section_classifier}. Each submission was annotated with its research hypothesis using prompts to GPT-3.5-turbo. To assess  annotation quality, we ask first authors to evaluate the annotations for their respective papers. More details on the annotation process and the results from the survey can be found in Appendix \ref{app:hypothesis_annotation}.

Additionally, we retrieved title and abstract of the references of all submissions and citation counts for accepted submissions from Semantic Scholar\footnote{\url{https://www.semanticscholar.org/}}. Summary statistics of the dataset are provided in Table \ref{tab:openreview_dataset_statistics} in Appendix \ref{app:openreview_dataset}. While all reviews are licensed under CC BY 4.0, not all submissions are accompanied by a license \footnote{\url{https://openreview.net/legal/terms}}. To ensure compliance, we publicly release the full dataset (excluding submissions) and provide access to submissions that have a CC BY 4.0 license. The complete dataset, including submissions, is accessible via a script from Huggingface. Details on the dataset creation process can be found in Appendix \ref{app:openreview_dataset}.

\textbf{ACL-OCL:} For the ACL-OCL dataset \cite{acl_ocl}, we update the citation scores, retrieve the title and abstract of the references and label the sections of the already provided GROBID parses and label each sample with its corresponding research hypothesis.

\section{Experiments}

First we study the relation between citation counts and review scores to understand how interchangeable these metrics are. Then we discuss the results for both score prediction tasks and compare the performance of simple score predictors with LLM based reviewers. Lastly, we provide a qualitative analysis to interpret the insights gained from the score prediction models.

\subsection{Citation Count vs. Review Scores}

We analyse the correlation between different review score dimensions and the correlation between citation counts and review scores. All review dimensions, except reviewer confidence, exhibit a positive correlation with the final score (see Figure \ref{fig:review_score_corr}). This indicates that reviewers, on average, do not exhibit varying levels of confidence when assigning high or low overall scores. Among the dimensions, impact and clarity show the strongest correlations with the final score. These results align partially with the findings of \citet{peer_read}, who analysed ACL 2017 reviews and identified clarity as a key factor, though their study reported a weaker correlation for impact compared to our findings.

The correlation between review scores and the logarithm of average citations per month across different venues is presented in Table \ref{tab:citation_review_score_prediction}. We analyze only venues from before 2024, allowing papers to have been published for at least one year as early citation counts exhibit high variance \cite{citation_prediction_sciencometrics}. Overall, we observe a weak positive correlation for most venues, with the exception of NeurIPS, where the correlation is negligible. For broad-topic conferences like ICLR and NeurIPS, a potential confounding factor is the varying popularity of different fields, which can influence citation counts regardless of the quality of the work. When separating ICLR-2023 papers by their author-provided field, we find mixed results: the correlation strengthens for some fields but disappears entirely for others (see Table \ref{tab:field_citation_correlation} in Appendix \ref{app:additional_results}).

\begin{table}[h!]
    \centering
    \begin{tabular}{l|c|c|}
     Dataset  & \# Samples & $\rho$  \\
     \hline 
     All & 15002 & 0.193   \\
     All-ICLR & 4920 & 0.148 \\
     ICLR - 2023 & 1507 & 0.168 \\
     ICLR - 2022 & 1072 & 0.175 \\
     ICLR - 2021 & 837 & 0.163 \\
     ICLR - 2020 & 674 & 0.120 \\
     ICLR - 2019 & 495 & 0.190\\
     ICLR - 2018 & 335 & 0.184\\ 
     NeurIPS - 2023 & 2963 & 0.103 \\
     NeurIPS - 2022 & 2553 & 0.085 \\
     NeurIPS - 2021 & 2285 & 0.085 
    \end{tabular}
    \caption{Pearson correlation between the log average citation per month and mean overall review scores for different subsets of the OpenReview dataset for accepted papers with at least one citation.}
    \label{tab:citation_review_score_prediction}
\end{table}

There only seems to be a weak overlap of the factors influencing citation count and review scores. Neither review scores nor citation counts alone can be considered definitive measures of scientific quality. Both metrics are influenced by external factors, such as author popularity or reviewer-field fit, that are unrelated to the intrinsic quality of the work. However, citation counts have the advantage of being easier to collect on a larger scale, making it easier to train models of larger scale which can be leveraged as an evaluation metric.

\subsection{Score Prediction}
\begin{table*}[t!]
    \centering
    \begin{tabular}{l|l|l|l|l|}
       \multirow{2}{*}{Context}  & \multicolumn{2}{c|}{Pairwise Comparison} & \multicolumn{2}{c|}{Regression} \\
        & Accuracy & $\rho_{s}$ & L1-Distance & $\rho_{s}$  \\
       \hline
        Title + Abstract & $\mathbf{0.665 (0.010)}$ & $\mathbf{0.481  (0.026)}$ & $\mathbf{0.921 (0.001)}$ & $\mathbf{0.498  (0.002)}$ \\
        Hypothesis & $0.615  (0.008)$ & $0.339  (0.024)$ & $ 1.003  (0.002)$  & $0.366  (0.001)$  \\
        Introduction & $0.655  (0.006) $ & $0.460  (0.016)$ & $0.943  (0.002)$ & $0.452  (0.002)$  \\
        Related Work & $0.631  (0.008)$ & $0.386 (0.022)$ & $0.955  (0.001)$ & $0.393  (0.002)$ \\
        Methodology & $0.634  (0.010) $ & $0.397  (0.027)$  & $0.981  (0.001)$ & $0.399  (0.002)$ \\
        Experiments \& Results & $0.651  (0.004)$ & $0.446  (0.012)$ & $0.946  (0.001)$ & $0.449  (0.001)$ \\
        Conclusion & $0.647 (0.005)$ &  $0.439  (0.014) $ & $0.944  (0.001)$ & $0.431  (0.002)$ \\
    
    \end{tabular}
    \caption{Performance of the citation prediction models for the pairwise comparisons and the prediction of the log average number of citation per month. All models are trained over five random seeds and mean results are presented with standard deviation in brackets. The best performing model is displayed in bold.}
    \label{tab:acl_ocl_citation_score_prediction}
\end{table*}

\textbf{Citation Score Prediction:}
In this section, we investigate the task of predicting the log average citations per month using various paper representations and contexts. Specifically, we conduct experiments to evaluate the predictive performance of different section types as paper representations, comparing them to representations based on \textit{Title + Abstract} and \textit{Hypothesis}. The results on the ACL-OCL dataset are presented in Table \ref{tab:acl_ocl_citation_score_prediction}. Both learning to rank and learning to predict exact citation scores achieve similar Spearman correlations with the ground-truth data. Contexts containing result-related information demonstrate higher prediction accuracy, with the most notable difference observed when comparing the \textit{Title + Abstract} context to the \textit{Hypothesis} context. Citation counts can still be predicted solely based on a paper's research hypothesis better than random. These results are in line with the intuition that empirical work will be cited more often if the results are more impressive. 

\begin{figure}
    \centering
    \includegraphics[width=\linewidth]{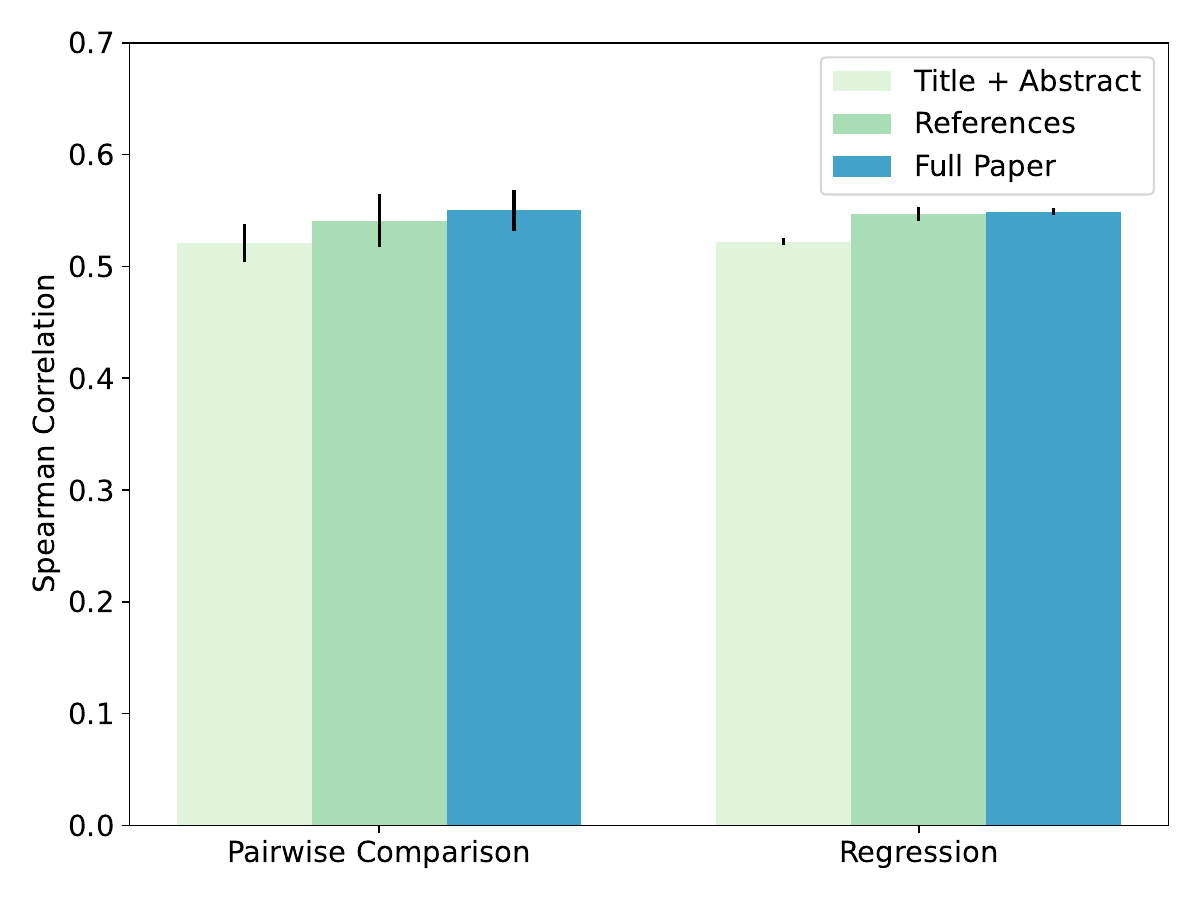}
    \caption{Spearman correlations for context-based models applied to both the regression and pairwise comparison tasks. Results are averaged over five seeds, with error bars representing the standard deviations.}
    \label{fig:acl_ocl_citation_score_prediction_context}
\end{figure}

Next, we train the context models and rerun the title and abstract baseline with an empty context to isolate the performance change caused by the architectural adjustment. The results show marginal benefits from training on complete paper information or adding the titles and abstracts of all references as context (see Figure \ref{fig:acl_ocl_citation_score_prediction_context}). A likely reason for the limited improvement is that most content-related variability in citation scores is already captured by the title and abstract. Remaining differences may stem from external factors, such as the authors’ networks. Additionally, the small dataset size constrains training to simpler models, as larger models tend to overfit and fail to generalize. Expanding datasets and utilizing full-text models could improve comparison accuracy in future studies.

\begin{table*}[]
    \centering
    \begin{tabular}{l|l|l|l|l|l|l}
       Dataset  &  Context & Metric &  \multicolumn{2}{c|}{Pairwise Comparison} & \multicolumn{2}{c}{Regression} \\
        & & & Accuracy & $\rho_{s}$ & L1-Distance &  $\rho_{s}$ \\
        \hline  
        
           \multirow{6}{*}{All}  & \multirow{3}{*}{TA} & Score & $0.506 (0.008)$ & $0.014 (0.025)$ & $0.094 (0.001)$ & $-0.006 (0.014)$ \\
                                                                     & & Impact & $0.541 (0.007)$ & $0.124 (0.018)$ &  $0.097 ( 0.005)$ & $0.109 (0.015)$  \\
                                                                     & & CC &  $0.608 (0.010)$ & $\mathbf{0.353 ( 0.031)}$ & $2.371 (0.013)$ & $0.330 (0.006)$\\
            & \multirow{3}{*}{H} & Score & $0.500 (0.006)$ & $0.001 (0.015)$ & $0.096 (0.005)$ & $0.010 (0.017)$ \\
                                          & & Impact & $0.535 ( 0.003)$ & $0.106 ( 0.008)$ & $0.102 ( 0.007)$ & $0.034 ( 0.042)$ \\
                                          & & CC & $0.590 ( 0.006)$ & $0.287 (0.015)$  & $2.236 (0.017)$ & $0.294 (0.007)$ \\
                                          
            \multirow{6}{*}{NeurIPS}  & \multirow{3}{*}{TA} & Score & $0.584 (0.004)$ & $0.252 (0.011)$ & $0.054 (0.001)$ & $0.242 (0.006)$   \\
                                                                     & & Impact & $0.542 (0.005)$ & $0.127 (0.013)$ & $0.091  (0.001)$ & $0.125 (0.014)$   \\
                                                                     & & CC &  $0.603 (0.007)$ & $\mathbf{0.322 (0.014)}$  & $2.132 (0.034)$ & $0.199 (0.017)$  \\
                                                                     
            & \multirow{3}{*}{H} & Score & $0.573 (0.004)$ & $0.216 (0.014)$  & $0.058 (0.002)$ & $0.160 (0.033)$     \\
                                          & & Impact & $0.528 (0.003)$ & $0.085 (0.011)$ & $0.093 (0.001)$ & $0.055 (0.014)$ \\
                                          & & CC&  $0.590 (0.005)$ & $0.288 (0.015)$  &  $2.089 (0.010)$ & $0.276 (0.009)$  \\

            \multirow{4}{*}{ICLR}  & \multirow{2}{*}{TA} & Score & $0.594 (0.004)$ & $0.274 (0.013)$ &  $0.117 (0.000)$ & $0.292 (0.006)$ \\
                                                                     & & CC &  $0.648 (0.001)$ & $0.433 (0.003)$ & $1.242 (0.005)$ & $\mathbf{0.443 (0.003)}$ \\
            & \multirow{2}{*}{H} & Score &  $0.571 (0.005)$ & $0.211 (0.017)$ &  $0.121 (0.000)$ & $0.196 (0.007)$  \\
                                          & & CC & $0.636 (0.002$) & $0.400 (0.005$) & $1.309 (0.007)$ & $0.399 (0.005$)\\

    \end{tabular}
    \caption{Prediction results on the OpenReview dataset for the different scores and subsets that the dataset contains. Models are run over five random seeds and the mean results are presented with standard deviation in brackets (TA=Title and Abstract, H=Hypothesis, CC=Citation Count). The best performing model based on the Spearman rank correlation for each subset of the dataset is indicated in bold.}
    \label{tab:openreview_prediction_results}
\end{table*}

\textbf{Review Score Prediction}: For review score prediction, our findings indicate that conference-agnostic prediction is challenging. On the full OpenReview dataset, the model's performance is no better than random guessing (see Table \ref{tab:openreview_prediction_results}). Two factors likely contribute to this difficulty. First, reviewing standards differ across conferences, meaning that even after normalization, the same paper could receive a score of 0.6 in one venue and 0.8 in another. Second, the broader variability in topics between conferences, as opposed to within a single conference, makes comparisons more difficult. Notably, venue-specific review score prediction models achieve a comparison accuracy of approximately $60\%$ for NeurIPS and ICLR.

A consistent trend across all dataset subsets is that predicting citation counts is easier than predicting review scores, resulting in higher comparison accuracies. As observed with the ACL-OCL dataset, predictions based on title and abstract perform better than those based on the research hypothesis. The comparison accuracy for the citation score is higher on the ACL-OCL dataset, one potential reason is the increased amount of data from a less broad range of topics. To check whether distinguishing by topics improves prediction performance on the ICLR subset we ran an latent dirichlet allocation (LDA) to separate all ICLR submissions into different topics. More details on the topic model can be found in Appendix \ref{app:appendix_lda}. We trained pairwise comparison score models for the five most frequent topics. From the results in Table \ref{tab:review_scores_topic_models}, it is evident that training separate prediction models per latent topic does not improve comparison accuracy.

\begin{table}[]
    \centering
    \begin{tabular}{l|l|l|}
         \# Samples &  \multicolumn{2}{c|}{Pairwise Comparison}  \\
         &  Accuracy & $\rho_{s}$  \\
         \hline
         5068 &  $0.580(0.002)$ & $0.236 (0.006)$   \\
         5542 &  $0.576(0.003)$ & $0.224(0.008)$   \\
         1624 &  $0.599(0.003)$ & $0.281(0.006)$  \\
         962 &  $0.565(0.006)$ & $0.182(0.016)$   \\
         707 &  $0.597 (0.002)$ & $0.285 (0.005)$  \\
    \end{tabular}
    \caption{Pairwise review score comparison accuracy for different topic subsets of the ICLR dataset, with topic labels for individual submissions generated using LDA.}
    \label{tab:review_scores_topic_models}
\end{table}

\subsection{Comparisons with LLM and Human Reviews}

\begin{figure*}[ht]
    \centering
    \subfigure[Sakana]{\includegraphics[width=0.32\textwidth]{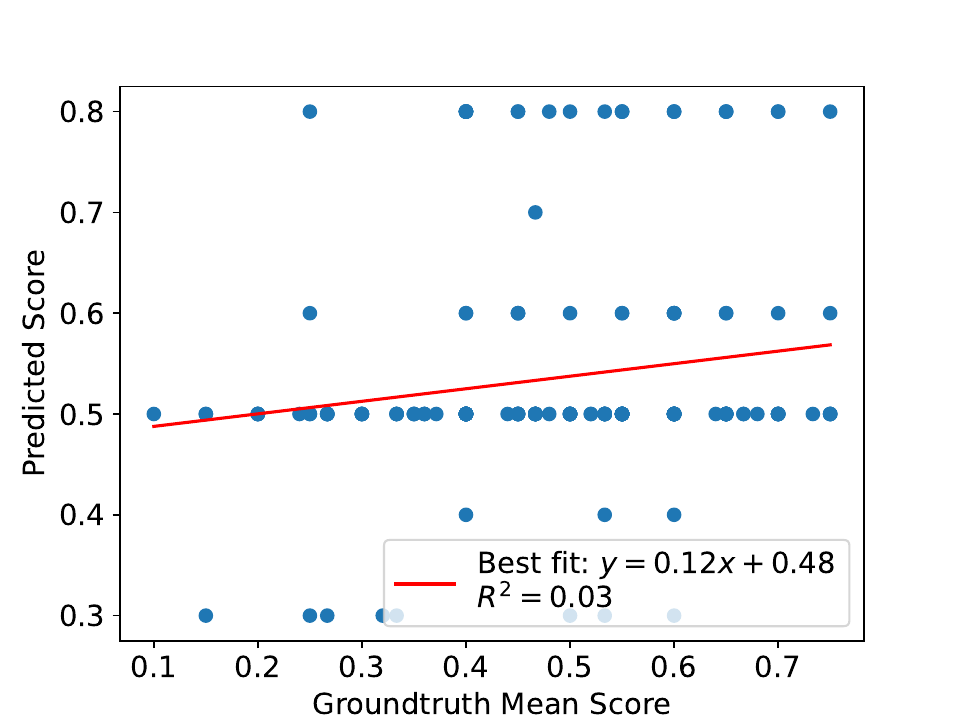}\label{fig:1}}
    \hfill
    \subfigure[RSP]{\includegraphics[width=0.32\textwidth]{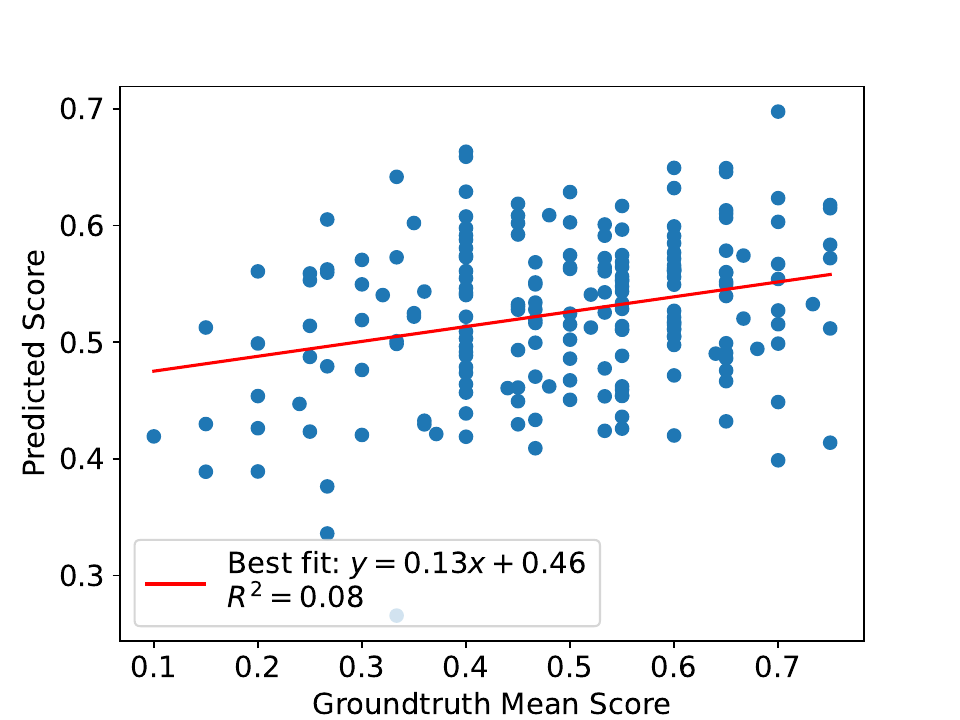}\label{fig:2}}
    \hfill
    \subfigure[Human]{\includegraphics[width=0.32\textwidth]{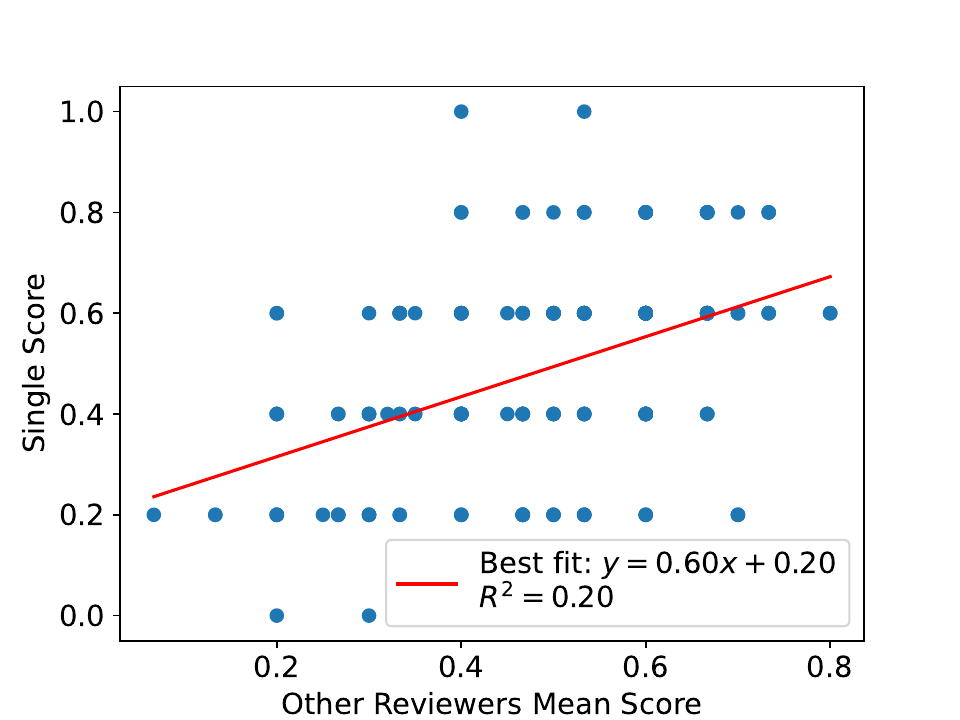}\label{fig:3}}
    \caption{Scatter plots of predicted review scores and groundtruth review scores on a subset of the test set of ICLR-2024 for the Sakana reviewer, the review score prediction model and human reviews. For human reviews, we randomly select a review as the predicted score and average over the rest.}
    \label{fig:scatter_plot_sakana}
\end{figure*}

To evaluate our review score prediction models against both LLM-based and human reviewers, we subsampled 200 papers from the test set of ICLR 2024 and NeurIPS 2024 respectively. We first tested the LLM-based reviewer from \citet{NLP100}, which takes two papers as input and predicts which one is more likely to receive a higher review score. Additionally, we apply our trained review score prediction models that take title and abstract as input to the same subsets.

As shown in Table \ref{tab:reviewer_pairwise_performance_comparison}, our simple prediction models, outperform the pairwise comparison accuracy of the LLM reviewer for both ICLR and NeurIPS. Two key differences exist compared to the original setup in \citet{NLP100}: (1) the LLM in the original work evaluated project proposals rather than full papers, and (2) the reviewed papers in the original setup were restricted to LLM-related topics, potentially simplifying comparisons. Despite these differences, the LLM-based reviewer demonstrates comparable performance to the original study. 

Further we compare the review score prediction model with the Sakana reviewer \cite{sakana}. In Figure \ref{fig:scatter_plot_sakana}, the relation between mean review score and predicted review scores are visualized. To compare against human reviewing performance, we exclude for each submission that has at least three reviews one review, and plot the relation between the single review score and the mean of the other review scores. The review score prediction model has a mean Pearson correlation over five random seeds of $0.330\pm 0.030$ which is higher than the correlation of the Sakana reviewer of $0.161$, but lower than the human reviewers correlation of $0.412\pm0.044$. Since results depend on which reviewer is sampled as a single reviewer, the human reviewer baseline is ran over five random seeds as well. Additionally, we can repeat the human reviewer consistency analysis for the whole ICLR and OpenReview dataset. The mean correlation is $0.504 \pm 0.004$ and $0.475\pm0.002$ respectively, again showing that human reviews show a higher consistency than LLM reviews with human reviews.

In Table \ref{tab:reviewer_comparison} (Appendix \ref{app:additional_results}), we qualitatively compare a review generated by the Sakana LLM reviewer with one written by a human reviewer for a ICLR-2024 submission. The LLM review highlights only generic strengths and weaknesses, while the human review provides detailed insights into the proposed methodology and its connections to the related work. This comparison illustrates that, despite a weak correlation in assigned scores, the LLM reviewer does not seem to base the scores on a deep semantic understanding of the paper.

\begin{table}[]
    \centering
    \begin{tabular}{l|l|l|}
        \multirow{2}{*}{Reviewer}  & \multicolumn{2}{c|}{Pairwise Comparison} \\
            & Accuracy & $\rho_{s}$  \\
         \hline
        LLM-ICLR & $0.548$ & $0.145$  \\
        RSP-ICLR   & $\mathbf{0.583 (0.012)}$ & $\mathbf{0.236 (0.050)}$ \\
        LLM-Neur. & $0.552$ & $0.064$  \\
        RSP-Neur. & $\mathbf{0.585 (0.013)}$ & $\mathbf{0.244 (0.034)}$ \\
    \end{tabular}
    \caption{Comparison accuracy and spearman correlation ($\rho_{s}$) for LLM-based and review-score comparison models on subsets of ICLR and NeurIPS test sets. Review score predictions are averaged over 5 random seeds, with results shown as mean (standard deviation).}
    \label{tab:reviewer_pairwise_performance_comparison}
\end{table}

\subsection{Qualitative Analysis}

To gain deeper insights into the features used by the citation and review score prediction models, we compute approximate Shapley values \cite{shap_values} for models trained on the ICLR subset of the dataset. Our analysis includes a comparison between review score and citation score prediction models, as well as an evaluation of models that use research hypothesis versus title and abstract as input. Visualizations of Shapley values for example inputs are provided in Appendix \ref{app:qualitative_analysis}. Comparing models using titles and abstracts to those based on research hypotheses shows that the former leverage result-oriented information. For example, phrases like "sub to superhuman performance" or "a speedup compared to state-of-the-art systems" consistently have high positive Shapley values. Models based on research hypotheses focus more on individual methods (e.g., NoisyNet, GAN) or topics (e.g., exploration in deep reinforcement learning, generative modeling). When comparing citation score and review score models, some phrases (e.g., "efficient exploration," "adversarial objectives") contribute oppositely to each score.
 
\section{Conclusion}

Our study highlights the potential of citation and review score prediction models as automatic evaluation metrics, demonstrating their alignment with human reviews and their improved performance compared to LLM-based review systems. However, human reviews are still more consistent than review score predictions with human review scores. The lack of a standardized review format across venues complicates comparisons across venues. Further, venues often only make submissions partially (NeurIPS) or not at all (ICML) available. Predicting citation scores is more viable than predicting review scores. Since content based citation prediction has not yet been deeply explored there are still questions to address related to scalability via larger datasets and the most effective target scores and input features. Moreover, excluding tables and figures potentially limits the potential of citation score prediction models \cite{cimate}.

\section*{Limitations}

While our score prediction models outperform LLM-based review systems, they still have limitations. Current models utilize only a small portion of the full paper's information, leaving room for improvement in incorporating full paper information and references. Limited dataset sizes may hinder current models' ability to leverage more information effectively. Furthermore, both citation count and review scores are imperfect proxies for scientific quality as they are influenced by factors such as the author's network, raising the open question of how accurately they can be predicted from paper content alone.

\section*{Acknowledgements}
This research was (partially) funded by the Hybrid Intelligence Center, a 10-year programme funded by the Dutch Ministry of Education, Culture and Science through the Netherlands Organisation for Scientific Research, https://hybrid-intelligence-centre.nl. This work used the Dutch national e-infrastructure with the support of the
SURF Cooperative using grant no. EINF-9756.

\bibliography{custom}

\begin{thebibliography}{51}
\providecommand{\natexlab}[1]{#1}

\bibitem[{Baek et~al.(2024)Baek, Jauhar, Cucerzan, and Hwang}]{research_agent}
Jinheon Baek, Sujay~Kumar Jauhar, Silviu Cucerzan, and Sung~Ju Hwang. 2024.
\newblock \href {https://doi.org/10.48550/ARXIV.2404.07738} {Researchagent: Iterative research idea generation over scientific literature with large language models}.
\newblock \emph{CoRR}, abs/2404.07738.

\bibitem[{Bai et~al.(2019)Bai, Zhang, and Lee}]{citation_metadata1}
Xiaomei Bai, Fuli Zhang, and Ivan Lee. 2019.
\newblock \href {https://doi.org/10.1016/J.JOI.2019.01.010} {Predicting the citations of scholarly paper}.
\newblock \emph{J. Informetrics}, 13(1):407--418.

\bibitem[{Bharti et~al.(2021)Bharti, Ranjan, Ghosal, Agrawal, and Ekbal}]{peer_assist}
Prabhat~Kumar Bharti, Shashi Ranjan, Tirthankar Ghosal, Mayank Agrawal, and Asif Ekbal. 2021.
\newblock \href {https://doi.org/10.1007/978-3-030-91669-5\_33} {Peerassist: Leveraging on paper-review interactions to predict peer review decisions}.
\newblock In \emph{Towards Open and Trustworthy Digital Societies - 23rd International Conference on Asia-Pacific Digital Libraries, {ICADL} 2021, Virtual Event, December 1-3, 2021, Proceedings}, volume 13133 of \emph{Lecture Notes in Computer Science}, pages 421--435. Springer.

\bibitem[{Bird(2006)}]{nltk}
Steven Bird. 2006.
\newblock \href {https://doi.org/10.3115/1225403.1225421} {{NLTK:} the natural language toolkit}.
\newblock In \emph{{ACL} 2006, 21st International Conference on Computational Linguistics and 44th Annual Meeting of the Association for Computational Linguistics, Proceedings of the Conference, Sydney, Australia, 17-21 July 2006}. The Association for Computer Linguistics.

\bibitem[{Blei et~al.(2003)Blei, Ng, and Jordan}]{lda_orig}
David~M. Blei, Andrew~Y. Ng, and Michael~I. Jordan. 2003.
\newblock Latent dirichlet allocation.
\newblock \emph{J. Mach. Learn. Res.}, 3(null):993–1022.

\bibitem[{Boiko et~al.(2023)Boiko, MacKnight, Kline, and Gomes}]{llm_chemical_research}
Daniil~A. Boiko, Robert MacKnight, Ben Kline, and Gabe Gomes. 2023.
\newblock \href {https://doi.org/10.1038/S41586-023-06792-0} {Autonomous chemical research with large language models}.
\newblock \emph{Nat.}, 624(7992):570--578.

\bibitem[{Bran et~al.(2024)Bran, Cox, Schilter, Baldassari, White, and Schwaller}]{chemistry_tools}
Andres~M. Bran, Sam Cox, Oliver Schilter, Carlo Baldassari, Andrew~D. White, and Philippe Schwaller. 2024.
\newblock \href {https://doi.org/10.1038/S42256-024-00832-8} {Augmenting large language models with chemistry tools}.
\newblock \emph{Nat. Mac. Intell.}, 6(5):525--535.

\bibitem[{D'Arcy et~al.(2024)D'Arcy, Hope, Birnbaum, and Downey}]{review_generation2}
Mike D'Arcy, Tom Hope, Larry Birnbaum, and Doug Downey. 2024.
\newblock \href {https://doi.org/10.48550/ARXIV.2401.04259} {{MARG:} multi-agent review generation for scientific papers}.
\newblock \emph{CoRR}, abs/2401.04259.

\bibitem[{de~Buy~Wenniger et~al.(2023)de~Buy~Wenniger, van Dongen, and Schomaker}]{multischubert}
Gideon~Maillette de~Buy~Wenniger, Thomas van Dongen, and Lambert Schomaker. 2023.
\newblock \href {https://doi.org/10.48550/ARXIV.2308.07971} {Multischubert: Effective multimodal fusion for scholarly document quality prediction}.
\newblock \emph{CoRR}, abs/2308.07971.

\bibitem[{Driess et~al.(2023)Driess, Xia, Sajjadi, Lynch, Chowdhery, Ichter, Wahid, Tompson, Vuong, Yu, Huang, Chebotar, Sermanet, Duckworth, Levine, Vanhoucke, Hausman, Toussaint, Greff, Zeng, Mordatch, and Florence}]{palm_e}
Danny Driess, Fei Xia, Mehdi S.~M. Sajjadi, Corey Lynch, Aakanksha Chowdhery, Brian Ichter, Ayzaan Wahid, Jonathan Tompson, Quan Vuong, Tianhe Yu, Wenlong Huang, Yevgen Chebotar, Pierre Sermanet, Daniel Duckworth, Sergey Levine, Vincent Vanhoucke, Karol Hausman, Marc Toussaint, Klaus Greff, Andy Zeng, Igor Mordatch, and Pete Florence. 2023.
\newblock \href {https://proceedings.mlr.press/v202/driess23a.html} {Palm-e: An embodied multimodal language model}.
\newblock In \emph{International Conference on Machine Learning, {ICML} 2023, 23-29 July 2023, Honolulu, Hawaii, {USA}}, volume 202 of \emph{Proceedings of Machine Learning Research}, pages 8469--8488. {PMLR}.

\bibitem[{Dycke et~al.(2023)Dycke, Kuznetsov, and Gurevych}]{other_dataset}
Nils Dycke, Ilia Kuznetsov, and Iryna Gurevych. 2023.
\newblock \href {https://doi.org/10.18653/V1/2023.ACL-LONG.277} {Nlpeer: {A} unified resource for the computational study of peer review}.
\newblock In \emph{Proceedings of the 61st Annual Meeting of the Association for Computational Linguistics (Volume 1: Long Papers), {ACL} 2023, Toronto, Canada, July 9-14, 2023}, pages 5049--5073. Association for Computational Linguistics.

\bibitem[{Esser et~al.(2024)Esser, Kulal, Blattmann, Entezari, M{\"{u}}ller, Saini, Levi, Lorenz, Sauer, Boesel, Podell, Dockhorn, English, and Rombach}]{stable_diffusion_3}
Patrick Esser, Sumith Kulal, Andreas Blattmann, Rahim Entezari, Jonas M{\"{u}}ller, Harry Saini, Yam Levi, Dominik Lorenz, Axel Sauer, Frederic Boesel, Dustin Podell, Tim Dockhorn, Zion English, and Robin Rombach. 2024.
\newblock \href {https://openreview.net/forum?id=FPnUhsQJ5B} {Scaling rectified flow transformers for high-resolution image synthesis}.
\newblock In \emph{Forty-first International Conference on Machine Learning, {ICML} 2024, Vienna, Austria, July 21-27, 2024}. OpenReview.net.

\bibitem[{Fernandes and de~Melo(2024)}]{automated_review_process}
Gustavo~L{\'{u}}cius Fernandes and Pedro O. S.~Vaz de~Melo. 2024.
\newblock \href {https://doi.org/10.1007/S00799-023-00382-1} {Enhancing the examination of obstacles in an automated peer review system}.
\newblock \emph{Int. J. Digit. Libr.}, 25(2):341--364.

\bibitem[{Fu and Aliferis(2008)}]{early_work_citation_prediction1}
Lawrence~D. Fu and Constantin~F. Aliferis. 2008.
\newblock \href {https://knowledge.amia.org/amia-55142-a2008a-1.625176/t-001-1.626020/f-001-1.626021/a-045-1.626429/a-046-1.626426} {Models for predicting and explaining citation count of biomedical articles}.
\newblock In \emph{{AMIA} 2008, American Medical Informatics Association Annual Symposium, Washington, DC, USA, November 8-12, 2008}. {AMIA}.

\bibitem[{Funkquist et~al.(2023)Funkquist, Kuznetsov, Hou, and Gurevych}]{writing_assistant1}
Martin Funkquist, Ilia Kuznetsov, Yufang Hou, and Iryna Gurevych. 2023.
\newblock \href {https://doi.org/10.18653/V1/2023.EMNLP-MAIN.455} {Citebench: {A} benchmark for scientific citation text generation}.
\newblock In \emph{Proceedings of the 2023 Conference on Empirical Methods in Natural Language Processing, {EMNLP} 2023, Singapore, December 6-10, 2023}, pages 7337--7353. Association for Computational Linguistics.

\bibitem[{Gao et~al.(2023)Gao, Yen, Yu, and Chen}]{writing_assistant2}
Tianyu Gao, Howard Yen, Jiatong Yu, and Danqi Chen. 2023.
\newblock \href {https://doi.org/10.18653/V1/2023.EMNLP-MAIN.398} {Enabling large language models to generate text with citations}.
\newblock In \emph{Proceedings of the 2023 Conference on Empirical Methods in Natural Language Processing, {EMNLP} 2023, Singapore, December 6-10, 2023}, pages 6465--6488. Association for Computational Linguistics.

\bibitem[{Hirako et~al.(2023)Hirako, Sasano, and Takeda}]{newly_published_citation_prediction}
Jun Hirako, Ryohei Sasano, and Koichi Takeda. 2023.
\newblock \href {https://doi.org/10.18653/V1/2023.FINDINGS-EACL.84} {Realistic citation count prediction task for newly published papers}.
\newblock In \emph{Findings of the Association for Computational Linguistics: {EACL} 2023, Dubrovnik, Croatia, May 2-6, 2023}, pages 1101--1111. Association for Computational Linguistics.

\bibitem[{Hirako et~al.(2024)Hirako, Sasano, and Takeda}]{cimate}
Jun Hirako, Ryohei Sasano, and Koichi Takeda. 2024.
\newblock \href {https://doi.org/10.48550/ARXIV.2410.04404} {Cimate: Citation count prediction effectively leveraging the main text}.
\newblock \emph{CoRR}, abs/2410.04404.

\bibitem[{Ib{\'{a}}{\~{n}}ez et~al.(2009)Ib{\'{a}}{\~{n}}ez, Larra{\~{n}}aga, and Bielza}]{early_work_citation_prediction2}
Alfonso Ib{\'{a}}{\~{n}}ez, Pedro Larra{\~{n}}aga, and Concha Bielza. 2009.
\newblock \href {https://doi.org/10.1093/BIOINFORMATICS/BTP585} {Predicting citation count of \emph{Bioinformatics} papers within four years of publication}.
\newblock \emph{Bioinform.}, 25(24):3303--3309.

\bibitem[{Jecmen et~al.(2023)Jecmen, Yoon, Conitzer, Shah, and Fang}]{malicious_bidding}
Steven Jecmen, Minji Yoon, Vincent Conitzer, Nihar~B. Shah, and Fei Fang. 2023.
\newblock \href {https://doi.org/10.1145/3543507.3583424} {A dataset on malicious paper bidding in peer review}.
\newblock In \emph{Proceedings of the {ACM} Web Conference 2023, {WWW} 2023, Austin, TX, USA, 30 April 2023 - 4 May 2023}, pages 3816--3826. {ACM}.

\bibitem[{Jr. et~al.(2024)Jr., Silva, Jr., and Amancio}]{embeddings_citation_count_prediction}
Adilson~Vital Jr., Filipi~N. Silva, Osvaldo N.~Oliveira Jr., and Diego~R. Amancio. 2024.
\newblock \href {https://doi.org/10.48550/ARXIV.2407.19942} {Predicting citation impact of research papers using {GPT} and other text embeddings}.
\newblock \emph{CoRR}, abs/2407.19942.

\bibitem[{Kang et~al.(2018)Kang, Ammar, Dalvi, van Zuylen, Kohlmeier, Hovy, and Schwartz}]{peer_read}
Dongyeop Kang, Waleed Ammar, Bhavana Dalvi, Madeleine van Zuylen, Sebastian Kohlmeier, Eduard~H. Hovy, and Roy Schwartz. 2018.
\newblock \href {https://doi.org/10.18653/V1/N18-1149} {A dataset of peer reviews (peerread): Collection, insights and {NLP} applications}.
\newblock In \emph{Proceedings of the 2018 Conference of the North American Chapter of the Association for Computational Linguistics: Human Language Technologies, {NAACL-HLT} 2018, New Orleans, Louisiana, USA, June 1-6, 2018, Volume 1 (Long Papers)}, pages 1647--1661. Association for Computational Linguistics.

\bibitem[{Lee et~al.(2022)Lee, Liang, and Yang}]{coauthor}
Mina Lee, Percy Liang, and Qian Yang. 2022.
\newblock \href {https://doi.org/10.1145/3491102.3502030} {Coauthor: Designing a human-ai collaborative writing dataset for exploring language model capabilities}.
\newblock In \emph{{CHI} '22: {CHI} Conference on Human Factors in Computing Systems, New Orleans, LA, USA, 29 April 2022 - 5 May 2022}, pages 388:1--388:19. {ACM}.

\bibitem[{Li et~al.(2024)Li, Patel, Wang, and Du}]{machine_learning_research_agent}
Ruochen Li, Teerth Patel, Qingyun Wang, and Xinya Du. 2024.
\newblock \href {https://doi.org/10.48550/ARXIV.2408.14033} {Mlr-copilot: Autonomous machine learning research based on large language models agents}.
\newblock \emph{CoRR}, abs/2408.14033.

\bibitem[{Li et~al.(2019)Li, Zhao, Yin, and Wen}]{citation_prediction_review_text1}
Siqing Li, Wayne~Xin Zhao, Eddy~Jing Yin, and Ji{-}Rong Wen. 2019.
\newblock \href {https://doi.org/10.18653/V1/D19-1497} {A neural citation count prediction model based on peer review text}.
\newblock In \emph{Proceedings of the 2019 Conference on Empirical Methods in Natural Language Processing and the 9th International Joint Conference on Natural Language Processing, {EMNLP-IJCNLP} 2019, Hong Kong, China, November 3-7, 2019}, pages 4913--4923. Association for Computational Linguistics.

\bibitem[{Liu and Shah(2023)}]{reviewer_gpt}
Ryan Liu and Nihar~B. Shah. 2023.
\newblock \href {https://doi.org/10.48550/ARXIV.2306.00622} {Reviewergpt? an exploratory study on using large language models for paper reviewing}.
\newblock \emph{CoRR}, abs/2306.00622.

\bibitem[{Lu et~al.(2024)Lu, Lu, Lange, Foerster, Clune, and Ha}]{sakana}
Chris Lu, Cong Lu, Robert~Tjarko Lange, Jakob Foerster, Jeff Clune, and David Ha. 2024.
\newblock \href {https://doi.org/10.48550/ARXIV.2408.06292} {The {AI} scientist: Towards fully automated open-ended scientific discovery}.
\newblock \emph{CoRR}, abs/2408.06292.

\bibitem[{Lundberg and Lee(2017)}]{shap_values}
Scott~M. Lundberg and Su{-}In Lee. 2017.
\newblock \href {https://proceedings.neurips.cc/paper/2017/hash/8a20a8621978632d76c43dfd28b67767-Abstract.html} {A unified approach to interpreting model predictions}.
\newblock In \emph{Advances in Neural Information Processing Systems 30: Annual Conference on Neural Information Processing Systems 2017, December 4-9, 2017, Long Beach, CA, {USA}}, pages 4765--4774.

\bibitem[{Mendoza et~al.(2022)Mendoza, Kusa, El{-}Ebshihy, Wu, Pride, Knoth, Herrmannova, Piroi, Pasi, and Hanbury}]{benchmark_scientific_document_classification}
{\'{O}}scar~E. Mendoza, Wojciech Kusa, Alaa El{-}Ebshihy, Ronin Wu, David Pride, Petr Knoth, Drahomira Herrmannova, Florina Piroi, Gabriella Pasi, and Allan Hanbury. 2022.
\newblock \href {https://aclanthology.org/2022.sdp-1.31} {Benchmark for research theme classification of scholarly documents}.
\newblock In \emph{Proceedings of the Third Workshop on Scholarly Document Processing, SDP@COLING 2022, Gyeongju, Republic of Korea, October 12 - 17, 2022}, pages 253--262. Association for Computational Linguistics.

\bibitem[{Muangkammuen et~al.(2022)Muangkammuen, Fukumoto, Li, and Suzuki}]{review_score_prediction_unlabeled}
Panitan Muangkammuen, Fumiyo Fukumoto, Jiyi Li, and Yoshimi Suzuki. 2022.
\newblock \href {https://doi.org/10.18653/V1/2022.FINDINGS-EMNLP.164} {Exploiting labeled and unlabeled data via transformer fine-tuning for peer-review score prediction}.
\newblock In \emph{Findings of the Association for Computational Linguistics: {EMNLP} 2022, Abu Dhabi, United Arab Emirates, December 7-11, 2022}, pages 2233--2240. Association for Computational Linguistics.

\bibitem[{Muangkammuen et~al.(2023)Muangkammuen, Fukumoto, Li, and Suzuki}]{review_score_prediction_intermediate}
Panitan Muangkammuen, Fumiyo Fukumoto, Jiyi Li, and Yoshimi Suzuki. 2023.
\newblock Intermediate-task transfer learning for peer review score prediction.
\newblock In \emph{Proceedings of the 13th International Joint Conference on Natural Language Processing and the 3rd Conference of the Asia-Pacific Chapter of the Association for Computational Linguistics: Student Research Workshop}, pages 40--47.

\bibitem[{Nair and Hinton(2010)}]{relu}
Vinod Nair and Geoffrey~E. Hinton. 2010.
\newblock \href {https://icml.cc/Conferences/2010/papers/432.pdf} {Rectified linear units improve restricted boltzmann machines}.
\newblock In \emph{Proceedings of the 27th International Conference on Machine Learning (ICML-10), June 21-24, 2010, Haifa, Israel}, pages 807--814. Omnipress.

\bibitem[{OpenAI(2023)}]{gpt4}
OpenAI. 2023.
\newblock \href {https://doi.org/10.48550/ARXIV.2303.08774} {{GPT-4} technical report}.
\newblock \emph{CoRR}, abs/2303.08774.

\bibitem[{Owen(1977)}]{owen_values}
Guilliermo Owen. 1977.
\newblock Values of games with a priori unions.
\newblock In \emph{Mathematical economics and game theory: Essays in honor of Oskar Morgenstern}, pages 76--88. Springer.

\bibitem[{Plank and van Dalen(2019)}]{citation_prediction_review_text2}
Barbara Plank and Reinard van Dalen. 2019.
\newblock \href {https://ceur-ws.org/Vol-2414/paper12.pdf} {Citetracked: {A} longitudinal dataset of peer reviews and citations}.
\newblock In \emph{Proceedings of the 4th Joint Workshop on Bibliometric-enhanced Information Retrieval and Natural Language Processing for Digital Libraries {(BIRNDL} 2019) co-located with the 42nd International {ACM} {SIGIR} Conference on Research and Development in Information Retrieval {(SIGIR} 2019), Paris, France, July 25, 2019}, volume 2414 of \emph{{CEUR} Workshop Proceedings}, pages 116--122. CEUR-WS.org.

\bibitem[{Qi et~al.(2023)Qi, Zhang, Li, Tian, Zeng, Chen, and Zhou}]{hypothesis_generation}
Biqing Qi, Kaiyan Zhang, Haoxiang Li, Kai Tian, Sihang Zeng, Zhang{-}Ren Chen, and Bowen Zhou. 2023.
\newblock \href {https://doi.org/10.48550/ARXIV.2311.05965} {Large language models are zero shot hypothesis proposers}.
\newblock \emph{CoRR}, abs/2311.05965.

\bibitem[{{\v R}eh{\r u}{\v r}ek and Sojka(2010)}]{gensim}
Radim {\v R}eh{\r u}{\v r}ek and Petr Sojka. 2010.
\newblock {Software Framework for Topic Modelling with Large Corpora}.
\newblock In \emph{{Proceedings of the LREC 2010 Workshop on New Challenges for NLP Frameworks}}, pages 45--50, Valletta, Malta. ELRA.
\newblock \url{http://is.muni.cz/publication/884893/en}.

\bibitem[{Rohatgi et~al.(2023)Rohatgi, Qin, Aw, Unnithan, and Kan}]{acl_ocl}
Shaurya Rohatgi, Yanxia Qin, Benjamin Aw, Niranjana Unnithan, and Min{-}Yen Kan. 2023.
\newblock \href {https://doi.org/10.18653/V1/2023.EMNLP-MAIN.640} {The {ACL} {OCL} corpus: Advancing open science in computational linguistics}.
\newblock In \emph{Proceedings of the 2023 Conference on Empirical Methods in Natural Language Processing, {EMNLP} 2023, Singapore, December 6-10, 2023}, pages 10348--10361. Association for Computational Linguistics.

\bibitem[{Si et~al.(2024)Si, Yang, and Hashimoto}]{NLP100}
Chenglei Si, Diyi Yang, and Tatsunori Hashimoto. 2024.
\newblock \href {https://doi.org/10.48550/ARXIV.2409.04109} {Can llms generate novel research ideas? {A} large-scale human study with 100+ {NLP} researchers}.
\newblock \emph{CoRR}, abs/2409.04109.

\bibitem[{Singh et~al.(2023)Singh, D'Arcy, Cohan, Downey, and Feldman}]{specter}
Amanpreet Singh, Mike D'Arcy, Arman Cohan, Doug Downey, and Sergey Feldman. 2023.
\newblock \href {https://doi.org/10.18653/V1/2023.EMNLP-MAIN.338} {Scirepeval: {A} multi-format benchmark for scientific document representations}.
\newblock In \emph{Proceedings of the 2023 Conference on Empirical Methods in Natural Language Processing, {EMNLP} 2023, Singapore, December 6-10, 2023}, pages 5548--5566. Association for Computational Linguistics.

\bibitem[{Srivastava et~al.(2014)Srivastava, Hinton, Krizhevsky, Sutskever, and Salakhutdinov}]{dropout}
Nitish Srivastava, Geoffrey~E. Hinton, Alex Krizhevsky, Ilya Sutskever, and Ruslan Salakhutdinov. 2014.
\newblock \href {https://doi.org/10.5555/2627435.2670313} {Dropout: a simple way to prevent neural networks from overfitting}.
\newblock \emph{J. Mach. Learn. Res.}, 15(1):1929--1958.

\bibitem[{Staudinger et~al.(2024)Staudinger, Kusa, Piroi, and Hanbury}]{review_dataset_overview}
Moritz Staudinger, Wojciech Kusa, Florina Piroi, and Allan Hanbury. 2024.
\newblock An analysis of tasks and datasets in peer reviewing.
\newblock In \emph{Proceedings of the Fourth Workshop on Scholarly Document Processing (SDP 2024)}, pages 257--268.

\bibitem[{van Dongen et~al.(2020)van Dongen, de~Buy~Wenniger, and Schomaker}]{schubert}
Thomas van Dongen, Gideon~Maillette de~Buy~Wenniger, and Lambert Schomaker. 2020.
\newblock \href {https://doi.org/10.18653/V1/2020.SDP-1.17} {Schubert: Scholarly document chunks with bert-encoding boost citation count prediction}.
\newblock In \emph{Proceedings of the First Workshop on Scholarly Document Processing, SDP@EMNLP 2020, Online, November 19, 2020}, pages 148--157. Association for Computational Linguistics.

\bibitem[{Vaswani et~al.(2017)Vaswani, Shazeer, Parmar, Uszkoreit, Jones, Gomez, Kaiser, and Polosukhin}]{attention}
Ashish Vaswani, Noam Shazeer, Niki Parmar, Jakob Uszkoreit, Llion Jones, Aidan~N. Gomez, Lukasz Kaiser, and Illia Polosukhin. 2017.
\newblock \href {https://proceedings.neurips.cc/paper/2017/hash/3f5ee243547dee91fbd053c1c4a845aa-Abstract.html} {Attention is all you need}.
\newblock In \emph{Advances in Neural Information Processing Systems 30: Annual Conference on Neural Information Processing Systems 2017, December 4-9, 2017, Long Beach, CA, {USA}}, pages 5998--6008.

\bibitem[{Wang et~al.(2021{\natexlab{a}})Wang, Peng, Zhang, and Zhang}]{findings_openreview}
Gang Wang, Qi~Peng, Yanfeng Zhang, and Mingyang Zhang. 2021{\natexlab{a}}.
\newblock \href {https://doi.org/10.1007/978-3-030-85896-4\_6} {What have we learned from openreview?}
\newblock In \emph{Web and Big Data - 5th International Joint Conference, APWeb-WAIM 2021, Guangzhou, China, August 23-25, 2021, Proceedings, Part {I}}, volume 12858 of \emph{Lecture Notes in Computer Science}, pages 63--79. Springer.

\bibitem[{Wang et~al.(2021{\natexlab{b}})Wang, Shi, Bai, Zhao, and Zhang}]{citation_history1}
Kehan Wang, Wenxuan Shi, Junsong Bai, Xiaoping Zhao, and Liying Zhang. 2021{\natexlab{b}}.
\newblock \href {https://doi.org/10.1007/S11192-021-04026-6} {Prediction and application of article potential citations based on nonlinear citation-forecasting combined model}.
\newblock \emph{Scientometrics}, 126(8):6533--6550.

\bibitem[{Wang et~al.(2024)Wang, Downey, Ji, and Hope}]{scimon}
Qingyun Wang, Doug Downey, Heng Ji, and Tom Hope. 2024.
\newblock \href {https://doi.org/10.18653/V1/2024.ACL-LONG.18} {Scimon: Scientific inspiration machines optimized for novelty}.
\newblock In \emph{Proceedings of the 62nd Annual Meeting of the Association for Computational Linguistics (Volume 1: Long Papers), {ACL} 2024, Bangkok, Thailand, August 11-16, 2024}, pages 279--299. Association for Computational Linguistics.

\bibitem[{Yang et~al.(2024)Yang, Du, Li, Zheng, Poria, and Cambria}]{hypothesis_generation2}
Zonglin Yang, Xinya Du, Junxian Li, Jie Zheng, Soujanya Poria, and Erik Cambria. 2024.
\newblock \href {https://doi.org/10.18653/V1/2024.FINDINGS-ACL.804} {Large language models for automated open-domain scientific hypotheses discovery}.
\newblock In \emph{Findings of the Association for Computational Linguistics, {ACL} 2024, Bangkok, Thailand and virtual meeting, August 11-16, 2024}, pages 13545--13565. Association for Computational Linguistics.

\bibitem[{Yuan et~al.(2022)Yuan, Liu, and Neubig}]{review_generation1}
Weizhe Yuan, Pengfei Liu, and Graham Neubig. 2022.
\newblock \href {https://doi.org/10.1613/JAIR.1.12862} {Can we automate scientific reviewing?}
\newblock \emph{J. Artif. Intell. Res.}, 75:171--212.

\bibitem[{Zhang and Wu(2024)}]{citation_prediction_sciencometrics}
Fang Zhang and Shengli Wu. 2024.
\newblock \href {https://doi.org/10.1007/S11192-024-05086-0} {Predicting citation impact of academic papers across research areas using multiple models and early citations}.
\newblock \emph{Scientometrics}, 129(7):4137--4166.

\bibitem[{Zhou et~al.(2024)Zhou, Chen, and Yu}]{review_generation3}
Ruiyang Zhou, Lu~Chen, and Kai Yu. 2024.
\newblock \href {https://aclanthology.org/2024.lrec-main.816} {Is {LLM} a reliable reviewer? {A} comprehensive evaluation of {LLM} on automatic paper reviewing tasks}.
\newblock In \emph{Proceedings of the 2024 Joint International Conference on Computational Linguistics, Language Resources and Evaluation, {LREC/COLING} 2024, 20-25 May, 2024, Torino, Italy}, pages 9340--9351. {ELRA} and {ICCL}.

\end{thebibliography}

\appendix

\newpage 

\section{Data Model} \label{app:data_model}

In the following we present the full data model for scientific papers, reviews and references. 

\begin{figure}[htb]
\centering

\begin{forest}
for tree={
  font=\small \ttfamily,
  grow'=0,
  child anchor=west,
  parent anchor=south,
  anchor=west,
  calign=first,
  edge path={
    \noexpand\path [draw, \forestoption{edge}]
    (!u.south west) +(1pt,0) |- node[fill,inner sep=1.25pt] {} (.child anchor)\forestoption{edge label};
  },
  before typesetting nodes={
    if n=1
      {insert before={[,phantom]}}
      {}
  },
  fit=band,
  before computing xy={l=15pt},
  s sep=1pt,
}
[Paper
  [Basic Information
      [Title]
      [Abstract]
      [Summary]
  ]
  [Parsed Content
      [Introduction]
      [Background]
      [Methodology]
      [Experiments and Results]
      [Conclusion]
  ]
  [Review Data
    [Decision]
    [Decision Text]
    [Reviews]
    [Comments]
  ]
  [IDs
    [Semantic Scholar Corpus ID]
    [ArXiv ID]
  ]
  [Metadata
    [Field of Study]
    [Number of Citations]
    [Number of Influential Citations]
  ]
  [References]
  [Research Hypothesis]
]
\end{forest}
\caption{Schematic overview of the scientific Paper object. The \textit{Field of Study} is a list of keywords that are part of the OpenReview submission, where the potential values depend on the venue. Number of citations and influential citations are retrieved from Semantic Scholar.}
\label{fig:paper_data_model}
\end{figure}
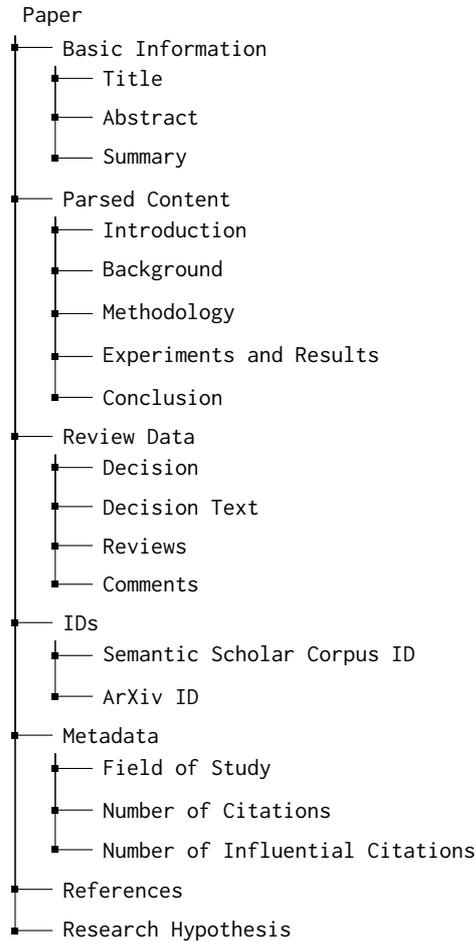

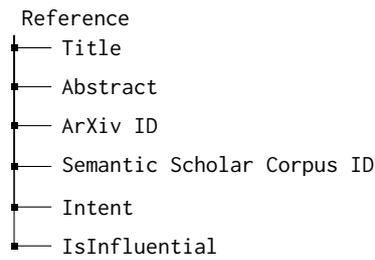
\begin{figure}
\begin{forest}
for tree={
  font=\small \ttfamily,
  grow'=0,
  child anchor=west,
  parent anchor=south,
  anchor=west,
  calign=first,
  edge path={
    \noexpand\path [draw, \forestoption{edge}]
    (!u.south west) +(1pt,0) |- node[fill,inner sep=1.25pt] {} (.child anchor)\forestoption{edge label};
  },
  before typesetting nodes={
    if n=1
      {insert before={[,phantom]}}
      {}
  },
  fit=band,
  before computing xy={l=15pt},
  s sep=1pt,
}
[Reference
[Title]
[Abstract]
[ArXiv ID]
[Semantic Scholar Corpus ID]
[Intent]
[IsInfluential]
]
\end{forest}
\caption{Schematic overview of the Reference object. The \textit{Intent} indicates the section of the scientific paper where the reference appears (e.g., introduction, methodology), while \textit{isInfluential} is a boolean value specifying whether the reference played a significant role in the creation of the paper. Both information come from Semantic Scholar.}

\label{fig:paper_object_forest}
\end{figure}

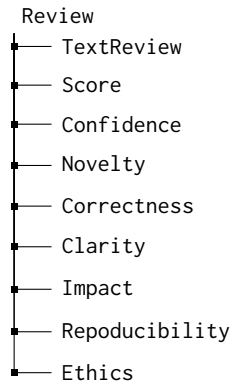
\begin{figure}
\begin{forest}
for tree={
  font=\small \ttfamily,
  grow'=0,
  child anchor=west,
  parent anchor=south,
  anchor=west,
  calign=first,
  edge path={
    \noexpand\path [draw, \forestoption{edge}]
    (!u.south west) +(1pt,0) |- node[fill,inner sep=1.25pt] {} (.child anchor)\forestoption{edge label};
  },
  before typesetting nodes={
    if n=1
      {insert before={[,phantom]}}
      {}
  },
  fit=band,
  before computing xy={l=15pt},
  s sep=1pt,
}
[Review
[TextReview]
[Score]
[Confidence]
[Novelty]
[Correctness]
[Clarity]
[Impact]
[Repoducibility]
[Ethics]
]
\end{forest}
\caption{Schematic overview of the review object. The \textit{TextReview} component concatenates all textual elements of the review, such as the summary, strengths, weaknesses, and questions. Except for the \textit{Ethics} text field, all other attributes of the review are represented as normalized floats ranging from 0 to 1.}
\label{fig:review_schema}
\end{figure}

\section{Training Details}\label{app:training_details}


\textbf{No-context model}: The no-context model consists of an MLP with 256 hidden units taking the SPECTER2 embeddings of size 768 as input. The training parameters are listed in Table \ref{tab:training_parameters_no_context}. We apply dropout of $0.3$ to the SPECTER2 embeddings. We conduct a grid search over learning rates $[0.0001,0.001,0.0005,0.00005]$ and dropout rates $[0,0.1,0.2,0.3,0.4,0.5]$, selecting the best-performing combination based on the validation set. Training is run for 100 epochs, and the checkpoint with the lowest validation loss is then evaluated on the test set.

\begin{table}[]
    \centering
    \begin{tabular}{l|l}
       Parameter  & Value  \\
       \hline
       Learning rate  &  0.00005 \\
       Dropout & 0.3 \\
       Epochs & 100 \\
       Batch Size & 256 \\
       Optimizer & Adam \\
       Hardware & \makecell[l]{NVIDIA GeForce \\ RTX 3080 (10GB)} \\
       Training Time & max. 20min \\       
    \end{tabular}
    \caption{Training parameters for the no context models for both citation and review score prediction on the OpenReview and ACL-OCL dataset.}
    \label{tab:training_parameters_no_context}
\end{table}

\textbf{Context Model}: The context model consists of a transformer encoder layer \cite{attention} with one head, ReLU activation \cite{relu}, dropout of $0.3$ and hidden units of size 1024. The paper representation and context is concatenated and processed via the encoder layer. The processed paper representation embedding is then passed to the same MLP as in the no-context model. Hyperparameter search and training are done in the same way as for the no-context model.

\begin{table}[]
    \centering
    \begin{tabular}{l|l}
       Parameter  & Value  \\
       \hline
       Learning rate  &  0.00005 \\
       Dropout & 0.3 \\
       Epochs & 50 \\
       Batch Size & 128 \\
       Optimizer & Adam \\ 
       Hardware & \makecell[l]{NVIDIA GeForce \\ RTX 3080 (10GB)} \\
       Training Time & max. 20min \\  
    \end{tabular}
    \caption{Training parameters for the context model.}
    \label{tab:training_parameters_context}
\end{table}

\section{OpenReview Dataset} \label{app:openreview_dataset}

\begin{figure}
    \centering
    \includegraphics[width=\linewidth]{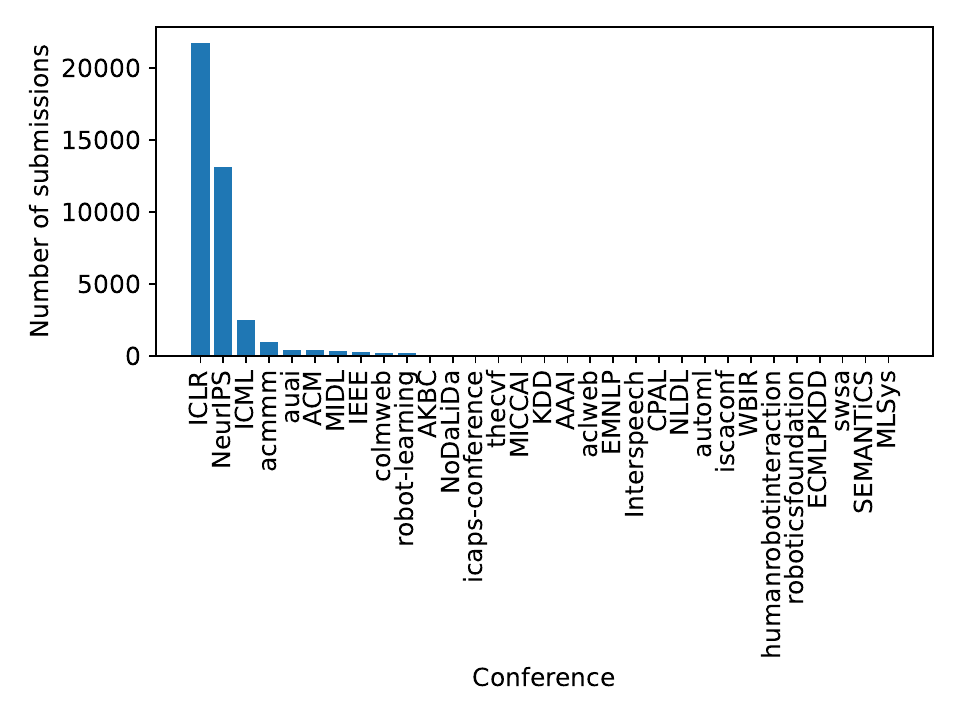}
    \caption{Number of submissions per conference}
    \label{fig:openreview_conference_distribution}
\end{figure}

In the following, we outline the collection process for the OpenReview dataset and present key dataset statistics. The dataset comprises submissions from OpenReview\footnote{\url{https://openreview.net/}} for which reviews and decisions are publicly accessible. The extent of accessibility varies by venue. For instance, ICML does not provide reviews or decisions, NeurIPS includes reviews only for accepted papers, and ICLR offers full access to all submissions. As illustrated in Figure \ref{fig:openreview_conference_distribution}, a significant proportion of the dataset originates from ICLR, followed by NeurIPS. For each venue, we manually map the review fields to the proposed review schema (see Table \ref{tab:review_field_harmonisation}) and extract both the decision and the decision text. The PDF document associated with each submission is downloaded and parsed using GROBID\footnote{\url{https://github.com/kermitt2/grobid}}. The parsed sections are then classified into categories from the paper schema (see Figure \ref{fig:paper_data_model}) using our section classifier (see Appendix \ref{app:section_classifier}). Additionally, we annotate each submission with a research hypothesis (see Appendix \ref{app:hypothesis_annotation}). For accepted submissions, we retrieve the citation count and influential citation count of the corresponding published papers from Semantic Scholar\footnote{\url{https://www.semanticscholar.org/product/api}}, along with their references. For rejected submissions, references are extracted from the parsed GROBID output and matched to entries in the Semantic Scholar corpus to obtain additional information. A summary of the dataset statistics is provided in Table \ref{tab:openreview_dataset_statistics}. OpenReview continues to grow in popularity, and we plan to update the dataset regularly to provide a growing dataset of papers with corresponding reviews in a unified format.

\begin{table}[]
    \centering
    \begin{tabular}{l|l}
       & Field   \\
       \hline
        & Recommendation\\
        & Confidence\\
        & Correctness\\
        & Empirical Novelty and Significance \\ 
        & Technical Novelty and Significance \\
        & Flag For Ethics Review: \\
        & Summary Of The Review: \\
        & Clarity, Quality, Novelty And Reproducibility \\
        & Strength And Weaknesses \\
        & Summary Of The Paper \\
    \end{tabular}
    \caption{Review fields for the ICLR-2023 venue.}
    \label{tab:iclr_2023_review_fields}
\end{table}

\onecolumn
\begin{longtable}{l|p{0.6\linewidth}} 
\hline
Review attribute & OpenReview review fields \\ \hline
\endfirsthead 
\hline
Review attribute & OpenReview review fields \\ \hline
\endhead 
\hline
\multicolumn{2}{r}{\textit{Continued on next page...}}  \\ 
\hline
\endfoot
\hline 
\caption{Harmonisation and mapping of all obtained OpenReview attributes to our review data model.} \label{tab:review_field_harmonisation}
\endlastfoot

score   & \makecell[l]{overall rating \\ rating \\ evaluation \\ Q6 Overall score \\ Overall score\\ recommended decision\\ Overall Score\\ overall evaluation\\ review rating \\ results\\ score\\ preliminary rating \\ recommendation \\ workshop rating \\ custom rating \\overall evaluation} \\ 
\hline
confidence & \makecell[l]{experience assessment \\ Reviewer expertise \\ confidence \\ review assessment: thoroughness in paper reading \\ reviewer's confidence\\ reviewer expertise\\ Confidence\\ Q8 Confidence in your score\\ review confidence \\ workshop confidence} \\
\hline
novelty & \makecell[l]{technical novelty and significance\\ originality\\ empirical novelty and significance\\ novelty\\ Q2(1) Originality/Novelty} \\
\hline
correctness & \makecell[l]{correctness\\ soundness\\ review assessment: checking correctness of experiments\\ Q2(3) Correctness/Technical quality\\ review assessment: checking correctness of derivations and theory\\ technical rigor\\ Q2(4) Quality of experiments (Optional)\\ technical quality and correctness rating\\ scholarship\\ technical quality\\ litreview } \\
\hline
clarity & \makecell[l]{presentation\\ clarity\\ clarity of presentation\\ Q2(6) Clarity of writing\\ clarity rating} \\
\hline
impact & \makecell[l]{ importance\\ contribution\\ relevance\\ impact\\ significance\\ Q2(2) Significance/Impact \\ potential impact on the field of AutoML rating \\significance and importance } \\ 
\hline
reproducibility & \makecell[l]{Q2(5) Reproducibility\\ reproducibility\\ usability and ease of reproducibility rating\\ accessibility} \\
\hline
paper summary & \makecell[l]{summary of the paper\\ summary of paper\\ problem statement\\ summary\\ Q1 Summary and contributions\\ summary of contributions} \\ 
\hline
review - summary & \makecell[l]{summary of recommendation\\ overall recommendation\\ summary of the review\\ reviewer confidence\\ Summary\\ justification of rating\\ Justification for rating\\ Q7 Justification for your score\\ review summary\\ overall reproducibility review} \\
\hline
main review & \makecell[l]{comment\\ intersection comment\\ detailed comments\\ rigor comment\\ clarity comment\\ main review\\ Q5 Detailed comments to the authors\\ potential impact on the field of AutoML\\ importance comment\\ technical quality and correctness\\ review\\ clarity\\ quality\\ novelty and reproducibility\\ issues\\ Q2 Assessment of the paper\\ overall review\\ usability and ease of reproducibility\\ review text\\ grounds for rejection\\ Main review \\ workshop review} \\ 
\hline
strength \& weakness & \makecell[l]{strengths weaknesses\\ strengths and weaknesses\\ weaknesses\\ Q3 Main strengths\\ strengths\\ strength and weaknesses\\ Q4 Main weakness\\ Review (Strengths/Weaknesses) \\Top Reasons to Accept the Paper \\ Top Reasons to Reject the Paper \\reason for not giving higher score \\ reason for not giving lower score \\ contributions of the paper \\ strengths of the paper \\ weaknesses of the paper \\ reasons to accept \\ reasons to reject} \\
\hline
limitations & \makecell[l]{limitations\\ limitations and societal impact\\ quality of the limitations section}\\
\hline
questions & \makecell[l]{questions for rebuttal\\ questions to address in the rebuttal\\ questions \\ Detailed Feedback and Questions for Authors \\ questions for authors} \\ 
\hline
ethics & \makecell[l]{ethics flag\\ Q10 Ethical concerns (Optional)\\ ethics and accessibility rating\\ ethics review area\\ flag for ethics review\\ details of ethics concerns\\ Ethical concerns\\ ethics details (optional)\\ needs ethics review\\ ethical concerns \\ ethical considerations } \\ 

\end{longtable}
\twocolumn

\begin{table}[]
    \centering
    \begin{tabular}{l|l}
        Statistic & Value  \\
        \hline
        \# Sub. with Reviews & 34646 \\
        \# Accepted Sub. & 22840 \\
        \# Workshop Sub. & 1817 \\
        \# Sub. with License & 6024 \\
        \# Sub. with Citation Count & 21314 \\
        \# Sub. with Introduction & 34043 \\
        \# Sub. with Related Work & 25826 \\
        \# Sub. with Methodology & 29850 \\
        \# Sub. with Experiments and Results & 32817 \\
        \# Sub. with Conclusion & 28035 \\
        \# Sub. with all sections & 19024 \\
        \# Reference Coverage Accepted Sub. & 86,4\% \\
        \# Reference Coverage Rejected Sub. & 87,3\% \\
        \# Rev. & 128602 \\
        \# Rev. with Clarity Score & 14773 \\
        \# Rev. with Confidence Score & 34263 \\
        \# Rev. with Correctness Score & 23962 \\
        \# Rev. with Reproducibility Score & 308 \\
        \# Rev. with Novelty Score & 6997 \\
        \# Rev. with Impact Score & 15172
    \end{tabular}
    \caption{Overview of the types of submissions (Sub.) and reviews (Rev.) and the availability of additional metadata.}
    \label{tab:openreview_dataset_statistics}
\end{table}

\section{Section Classification}\label{app:section_classifier}

Academic papers typically follow a standardized structure. To enable flexible use of different sections of a paper as input for our prediction models, we aim to map each of them to one of the section types outlined in the Paper Data Model (Figure \ref{fig:paper_data_model}). This requires training a section classifier. To construct the necessary dataset, we first define synonyms for section types commonly found as headings in academic papers (see Table \ref{tab:synonyms_section_headings}). We then process the OpenReview dataset and ACL-OCL dataset, identifying sections where the heading matches one of the predefined synonyms. These sections are subsequently added to the section classification dataset along with their corresponding labels\footnote{\url{https://huggingface.co/datasets/nhop/academic-section-classification}}. An overview of the frequency of the different section types in the two datasets can be found in Table \ref{tab:section_classification_dataset_frequency}. 

The section classifier takes a paragraph of text as input and assigns one of the five section types from the paper data model to it. First, the paragraph is split into sentences with NLTK's sentence tokenizer \cite{nltk}. Sentence embeddings are obtained via Specter2 by averaging over all token embeddings for the sentence. The sequence of sentence embeddings are then processed via two transformer encoder layers with dropout $0.3$, hidden dimension of 1024 and eight heads. The processed embeddings are then averaged and passed through a linear layer to obtain logits. The model is trained by minimizing the cross entropy loss. The dataset is split into a training, validation and test set where the training set makes up $70\%$ of the dataset and the validation and test set $15\%$ each. In Table \ref{tab:section_classifier_training_details} one can find relevant training parameters. The accuracy over 4 random seeds of the classifier on the test set is $0.921 \pm 0.006$ for the ACL-OCL dataset and $0.93 \pm 0.1$ for the OpenReview dataset.

\begin{table}[]
    \centering
    \begin{tabular}{l|l}
        Parameter & Value  \\
        \hline
        Learning Rate & 0.0001 \\
        Batch Size & 128 \\
        \# Epochs & 20 \\
        Hardware & \makecell[l]{NVIDIA GeForce \\ RTX 3080 (10GB)} \\
        Training Time & 25min
    \end{tabular}
    \caption{Training parameters for the section classifier training.}
    \label{tab:section_classifier_training_details}
\end{table}

\begin{table}[]
    \centering
    \begin{tabular}{l|l}
        Section Type & Synonyms  \\
        \hline
        Introduction & Introduction \\
         & \\
        Background & \makecell[l]{Background \\ Related Work \\ Historical Review} \\
         & \\
        Methodology & \makecell[l]{Methodology \\ Method \\ Algorithm \\ Properties} \\
         & \\
        E\&R & \makecell[l]{ Experiments \\ Results \\ Experiments and Results \\ Experimental Design \\ Empirical Evaluation \\ Experiments and Analysis \\ Ablation Studies \\ Evaluation} \\
         & \\
        Conclusion & \makecell[l]{Conclusion \\
        Conclusion \& Discussion \\
        Discussion and Conclusions \\
        Conclusion and Outlook \\
        Further Work \\
        Discussions and \\ Future Directions}\\
        
    \end{tabular}
    \caption{Section synonyms used to collect a training dataset for the section classifier (E\&R = Experiments and Results).}
    \label{tab:synonyms_section_headings}
\end{table}

\begin{table}[]
    \centering
    \begin{tabular}{l|c|c}
         Sections & \# OpenReview & \# ACL-OCL \\
         \hline
          Introduction & 39487 & 56223 \\
          Background & 20521 & 23412 \\
          Methodology & 6656 & 4302 \\
          E\&R & 18801 &  28157 \\
          Conclusion & 22811 & 28157 \\ 
    \end{tabular}
    \caption{Number of samples in the section classification dataset resulting from matching section headings of the GROBID PDF parses to the section type synonyms for the OpenReview and ACL-OCL dataset (E\&R=Experiments and Results).}
    \label{tab:section_classification_dataset_frequency}
\end{table}

\section{Research Hypothesis Annotation}\label{app:hypothesis_annotation}

To understand whether it is possible to predict citation and review scores based on the research hypothesis of a paper alone, we annotate the papers in the ACL-OCL and OpenReview dataset using \textit{gpt-3.5-turbo}. Following \citet{research_agent} we model a research hypothesis $h=[p,s]$ as a problem and a methodology/solution to solve that problem represented via natural language. Unlike \citet{research_agent} we do not include the experimental design as part of the research hypothesis. The one-shot prompt used to query the LLM can be seen in Table \ref{tab:annotation_prompt}. No more advanced LLM is used for annotation to restrict costs, as the context windows are large due to the content of the scientific papers. The number of annotation examples is restricted to 1 to reduce the context size, as the context window of \textit{gpt-3.5-turbo} is restricted to 16000 tokens.


\begin{table*}[tp]
\label{app:prompt_hypothesis_annotation}
\begin{tabular}{l p{0.8\linewidth}}
\hline
Types          & Texts                      \\ \hline
System Message &  Task Description: \newline
        You are a PhD student tasked to annotate research papers with the hypothesis they investigate.
        You will be provided with infos about the paper.
        Your task is to extract the research hypothesis from this provided text. \newline
        Requirements: \newline
        - Clarity: Ensure the hypothesis is clearly stated and understandable without additional context. \newline
        - Completeness: The hypothesis should be self-contained, including all necessary components such as the variables involved and the expected relationship or outcome. \newline
        - Terminology: Use precise and field-specific terminology that a research scientist in the relevant or adjacent field would understand. \newline
        - Conciseness: Keep the hypothesis one to two sentences long, avoiding unnecessary details or jargon. \newline
        Examples:

        Paper 1:
        \textbf{\{example\_paper\_text\}}

        Hypothesis: \textbf{\{example\_hypothesis\}}
        
        Format: \newline
        - Problem: The problem that the paper is addressing \newline
        - Solution: The solution that the paper is proposing \newline                                                                                              \\
        \hline
User Message   &  Annotate the following paper with its hypothesis: \textbf{\{paper\_text\}} \\ \hline
\end{tabular}
\caption{Few-shot prompt template for research hypothesis annotation.}
\label{tab:annotation_prompt}
\end{table*}

\begin{figure*}
    \centering
    \includegraphics[width=\linewidth]{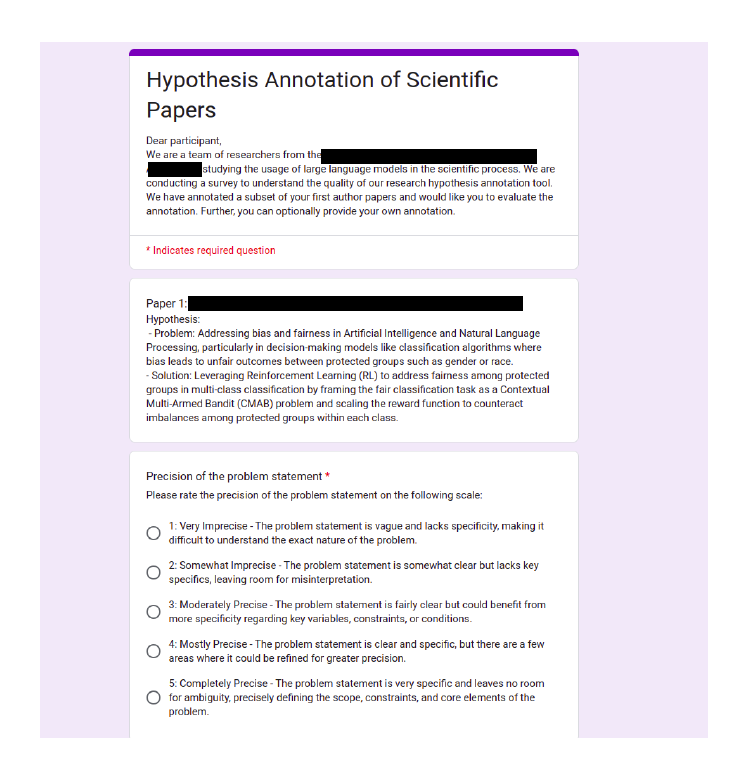}
    \caption{Example of the Google Form used to collect the survey data. Parts of the form is blacked out to guarantee annonymity.}
    \label{fig:google_form}
\end{figure*}

To verify that our research hypothesis annotation tool does indeed capture the main ideas of the papers, we asked first authors of academic papers to rate the quality of the annotated research hypothesis. The annotation quality is rated on the dimensions of correctness, precision and completeness for the problem as well as the solution. The 5 point Likert scales for each dimension is displayed in Table \ref{tab:likert_scales}. The figure \ref{fig:google_form} shows the Google Form used to conduct the survey. The survey was filled out by 13 first authors (PhD students and assistant professors) rating a total of 32 research hypotheses. The results are displayed in Table \ref{tab:survey_results}. The results show that the correctness of the annotated research hypotheses is overall satisfactory, however the completeness can be improved. In Table \ref{tab:annotation_examples}, some examples of the research hypothesis and improved versions written by first authors can be found. It is important to note that a potential source of error comes from mistakes in GROBID PDF parsing step that is used to extract the paper text used to condition the paper text.

\onecolumn
\begin{longtable}{l|p{0.7\linewidth}} 
\hline
Dimension & 5-point Likert scale \\ \hline
\endfirsthead 
\hline
Dimension & 5-point Likert scale  \\ \hline
\endhead 
\hline
\multicolumn{2}{r}{\textit{Continued on next page...}} \\ 
\hline
\endfoot
\hline 
\caption{The 5-point Likert scales used to evaluate the quality of each component of the annotated research hypothesis (PS = Problem Statement, S.=Solution, Compl.=Completeness, Correct.=Correctness).} \label{tab:likert_scales}
\endlastfoot

PS Correct.  & 1: The problem statement is fundamentally incorrect or misrepresents the actual problem or context. \newline
2: The problem statement has significant errors or misconceptions about the nature of the problem but some aspects are correct \newline
3: The problem statement is mostly correct but contains a few incorrect or misleading details about the nature or scope of the problem. \newline
4: The problem statement is almost entirely correct, with only minor inaccuracies or misinterpretations. \newline
5: The problem statement is entirely accurate, with no errors or misrepresentations regarding the nature or scope of the problem. \\
\hline
PS Precision   &  1: The problem statement is vague and lacks specificity, making it difficult to understand the exact nature of the problem. \newline 
2: The problem statement is somewhat clear but lacks key specifics, leaving room for misinterpretation. \newline 
3: The problem statement is fairly clear but could benefit from more specificity regarding key variables, constraints, or conditions. \newline 
4: The problem statement is clear and specific, but there are a few areas where it could be refined for greater precision. \newline 
5: The problem statement is very specific and leaves no room for ambiguity, precisely defining the scope, constraints, and core elements of the problem. \\
\hline
PS Compl.   & 1: The problem statement is vague, missing core details, and does not define the central challenge or its scope. \newline
     2: The problem statement captures some aspects of the challenge but omits critical elements such as specific requirements, constraints, or the broader context.\newline
     3: The problem statement conveys the general challenge but lacks clarity on important technical aspects, such as the specific domain or sub-problems within the larger issue.\newline
     4: The problem statement is thorough and covers most key aspects of the challenge, including domain, scope, and constraints, with only minor details or nuances missing.\newline
     5: The problem statement is fully articulated, covering all relevant aspects, including scope, context, constraints, and implications, leaving no critical information out. \\ 
     & \\
    \hline
    S. Correct.   &  1: The solution statement is fundamentally incorrect or misrepresents how the problem is addressed or resolved. \newline
       2: The solution statement has significant errors or misconceptions about how the problem is solved, but some aspects of the approach are correct. \newline
       3: The solution statement is mostly correct but contains a few incorrect or misleading details about the approach or method used to solve the problem. \newline
       4: The solution statement is almost entirely correct, with only minor inaccuracies or misinterpretations regarding the proposed solution. \newline
       5: The solution statement is entirely accurate, with no errors or misrepresentations regarding the approach, methods, or steps taken to solve the problem. \\
     & \\
     \hline
    S. Precision   &  1: The solution statement is vague and lacks specificity, making it difficult to understand the proposed approach or how it addresses the problem.\newline
      2: The solution statement is somewhat clear but lacks key specifics, leaving room for misinterpretation about how the solution addresses the problem.\newline
      3: The solution statement is fairly clear but could benefit from more specificity regarding key methods, steps, or conditions that explain how the problem is being solved.\newline
      4: The solution statement is clear and specific, but there are a few areas where it could be refined for greater precision or clarity regarding the approach.\newline
      5: The solution statement is very specific and leaves no room for ambiguity, precisely defining the method, scope, and key elements of the proposed solution. \\
     & \\
     \hline
    S. Compl.  &  1: The solution statement is vague, missing core details, and does not explain how the problem is addressed or resolved. \newline
       2: The solution statement captures some aspects of the approach but omits critical elements such as the specific methods, techniques, or constraints. \newline
       3: The solution statement explains the general approach but lacks clarity on important technical aspects, such as specific steps, tools, or techniques needed to solve the problem. \newline
       4: The solution statement is thorough and covers most key aspects of the approach, including methods, scope, and constraints, with only minor details missing. \newline
       5: The solution statement is fully articulated, covering all relevant aspects, including the method, tools, constraints, and implications, leaving no critical information out. \\

\end{longtable}
\twocolumn

\begin{table}[]
    \centering
    \begin{tabular}{l|l}
        Criteria & Rating  \\
        \hline
        Problem Correctness & $4.125  (1.083)$ \\
        Problem Precision & $3.688  (1.158) $  \\
        Problem Completeness & $3.563  (1.144)$ \\
        Solution Correctness & $4.125  (1.293)$ \\
        Solution Precision & $3.906  (1.042) $ \\
        Solution Completeness & $3.625  (1.192)$ \\
    \end{tabular}
    \caption{Average ratings of the annotated research hypotheses on the dimensions of correctness, precision, and completeness, measured using a 5-point Likert scale. Standard deviations are provided in parentheses.}
    \label{tab:survey_results}
\end{table}

\begin{table*}[tp]
\label{tab:h-example}
\begin{tabular}{l p{0.8\linewidth}}
\hline
Examples         & Texts                      \\ \hline
Example 1 &  
\textbf{LLM (gpt-3.5.-turbo)}:
\newline
\newline
\textbf{Problem:} The paper aims to investigate how conventions develop in task-based interactions, specifically focusing on how common ground influences the formation of conventions in the presence of new information, such as outer circle characters. \textit{(Precision:4, Correctness:5, Completeness:4)}

\textbf{Solution}: The paper proposes the SPOTTER framework, a gameplay framework for task-based interaction, to study how referring expressions to known 'inner circle' (InC) referents and unknown 'outer circle' (OutC) referents evolve over time, aiming to understand how conventions arise and how they are influenced by changing contexts. \textit{(Precision:5, Correctness:5, Completeness:5)}
\newline
\newline
\textbf{First Author Correction}:
\newline
\newline
\textbf{Problem:} The paper aims to investigate how conventions develop in task-based interactions \textcolor{red}{in Human-Robot Interaction}, specifically focusing on how common ground influences the formation of conventions in the presence of new information, such as outer circle characters.

\textbf{Solution:} The paper proposes the SPOTTER framework, a gameplay framework for task-based interaction, to study how referring expressions to known 'inner circle' (InC) referents and unknown 'outer circle' (OutC) referents evolve over time, aiming to understand how conventions arise and how they are influenced by changing contexts.
\newline
\\
\hline
Example 2  &  
\textbf{LLM (gpt-3.5.-turbo)}:
\newline
\newline
\textbf{Problem:} 
The paper aims to investigate the impact of common ground in social dialogues on the resolution of co-reference, specifically focusing on the distinction between "inner circle" and "outer circle" references in conversations. \textit{(Precision:4, Correctness:4, Completeness:3)}

\textbf{Solution:}
The hypothesis is that resolving references to well-known "inner circle" individuals is more challenging compared to lesser-known "outer circle" individuals in social dialogues, and that training models on preceding data may not effectively acquire common ground knowledge for inner circle references but can improve performance for outer circle mentions. \textit{(Precision:3, Correctness:2, Completeness:3)}
\newline

\textbf{First Author Correction}:
\newline
\newline
\textbf{Problem:} The paper aims to investigate the impact of common ground in social dialogues on co-reference resolution, specifically focusing on the distinction between "inner circle" and "outer circle" references in conversations.

\textbf{Solution:} The hypothesis is that resolving references to well-known "inner circle" individuals is more challenging compared to lesser-known "outer circle" individuals in social dialogues, and that training models on preceding data may help in acquiring common ground knowledge for inner circle references. \textcolor{red}{To test this, a data set of social dialogue is analysed on referring expressions for 'inner circle' and 'outer circle', and a co-reference resolution model is trained on preceding data and its performance analysed.}
\newline\\
\hline
\end{tabular}
\caption{Examples of annotated research hypotheses, their ratings and the corrected versions by the corresponding first authors. The highlighted text in red describes the parts that is changed by the first author}
\label{tab:annotation_examples}
\end{table*}

\section{Additional Results}\label{app:additional_results}

Here, we present additional resutls. In Table \ref{tab:field_citation_correlation} one can see the correlation between citation count and review score for different topics of the venue ICLR 2023. 

\begin{table}[]
    \centering
    \begin{tabular}{l|c|c}
        Field of Study & \# Samples & $\rho$  \\
        \hline
        All & 1507 & 0.168 \\
        \makecell[l]{Applications}& 186 & 0.231\\
        \makecell[l]{Deep Learning and \\ representational learning} & 342 & 0.236 \\
        \makecell[l]{General Machine Learning} & 57 & 0.190 \\
        \makecell[l]{Generative models}  & 63 & 0.210 \\
        \makecell[l]{ML for Sciences}& 59 & -0.135 \\
        Reinforcement Learning & 138 & 0.138 \\
        Social Aspects of ML & 93 & 0.198 \\
        Theory & 93 & 0.125 \\
        \makecell[l]{Unsupervised and \\ Self-supervised learning} & 71 & -0.009
        
    \end{tabular}
    \caption{Pearson correlation between the log average citation per month and the mean overall review score for accepted papers with at least one citation for the ICLR-2023 venue for the different fields of study. Fields of study with less than 50 entries are left out.}
    \label{tab:field_citation_correlation}
\end{table}

\onecolumn
\begin{longtable}{l|p{0.3\linewidth}|p{0.5\linewidth}} 
\hline
Review Field & Sakana Review & Human Review \\ \hline
\endfirsthead 
\hline
Review Field & Sakana Review & Human Review \\ \hline
\endhead 
\hline
\multicolumn{3}{r}{\textit{Continued on next page...}} \\ 
\hline
\endfoot
\hline 
\caption{Comparison of human review and review produced by Sakana on ICLR-2024 submission (S.=Soundness,P.=Presentation,C.=Correctness,CF.=Confidence).} \label{tab:reviewer_comparison}
\endlastfoot

Summary & The paper introduces a Fair Mapping method to mitigate bias in text-guided diffusion models, focusing on generating human-related images. The method is model-agnostic and lightweight, using a linear mapping network to address biases. A novel fairness evaluation metric is proposed, and experiments demonstrate the method's effectiveness. & This paper addresses human-related bias in text-to-image diffusion models, and resolves the issue by proposing a novel fair mapping module which outputs a fair text embedding. Such a module can be trained on top of frozen pre-trained text encoder, and inserting the module during sampling successfully mitigates textual bias. Training the fair module involves two loss terms: (i) text consistency loss, which preserves semantic coherence, and (ii) fair distance penalty, which brings output embeddings within different sensitive groups close together. Further, the authors propose a novel evaluation metric, FairScore, which also plans to achieve the conditional independence of the text prompt and sensitive group information.   \\
Weaknesses & Moderate originality as it builds on existing concepts. \newline
             Experimental evaluation is limited and lacks comprehensive comparison with state-of-the-art methods. \newline
             Insufficient analysis of computational complexity and scalability. \newline
             Certain sections, particularly the methodology, could be clearer.  & Although the paper covers a good amount of relevant previous studies, the paper lacks baseline experiments. For example, despite [1] focus on fair-guidance while this work focus on pluggable mapping module, the authors can calculate FairScore and compare w.r.t. training time, overhead memory, etc. \newline
            While the unfairness is largely resolved through the proposed mapping module, such a result may not come at a surprise since FairScore and the employed fairness loss term are quite similar. \newline
            The authors note that a detector network is employed to identify predefined sensitive keywords in the input prompts. There is no additional detailed explanation about the detector network. \newline
            This method explicitly needs a labeled dataset to mitigate the demographic bias in diffusion models. However, in real-world scenarios, it may be challenging to identify and address all potential types of bias comprehensively. Further, there are remaining questions regarding whether it is feasible to (i) simultaneously eliminate multiple types of bias or (ii) sequentially address multiple biases without negatively impacting performance. If such challenges cannot be properly addressed, it would incur a significant amount of training time to erase all types of biases, and heavy memory cost to save all mapping modules corresponding to each bias type. \\
Strength &  Addresses the critical issue of bias in text-guided diffusion models. \newline
            Proposes a novel Fair Mapping method that is model-agnostic and lightweight. \newline
            Introduces a new fairness evaluation metric, contributing to the field of bias assessment. \newline & 
            The paper tackles a timely and practically-relevant problem supported by a fair amount of experiments. Building  fair diffusion models is an area with limited prior research, making this work particularly valuable. \newline
            The proposed method is simple yet effective, and pluggable without modifying the pre-trained model. \newline
            Overall, the paper is clearly written and easy to follow. \newline \\
Questions &  How does the Fair Mapping method quantitatively compare with other state-of-the-art bias mitigation techniques? \newline
             What are the computational complexity and scalability considerations for the Fair Mapping method? \newline 
             Could the authors provide more implementation details, particularly regarding the training of the Fair Mapping network? 
              & How many random seeds are used throughout the experiments?  \\
Limitations & The method may not fully eliminate biases due to intertwined information from diverse modalities. \newline  Evaluation is limited by the lack of comparison with state-of-the-art methods. & \\
Ethical C. & False & No ethics review needed. \\
S. /P. / C. / CF.  & 3 / 3 / 3 / 4  & 3 / 3 / 2 / 4   \\
Overall & 5  & 5 \\
\hline
\end{longtable}

\twocolumn

\section{Topic Model}\label{app:appendix_lda}

\begin{figure}
    \centering
    \includegraphics[width=\linewidth]{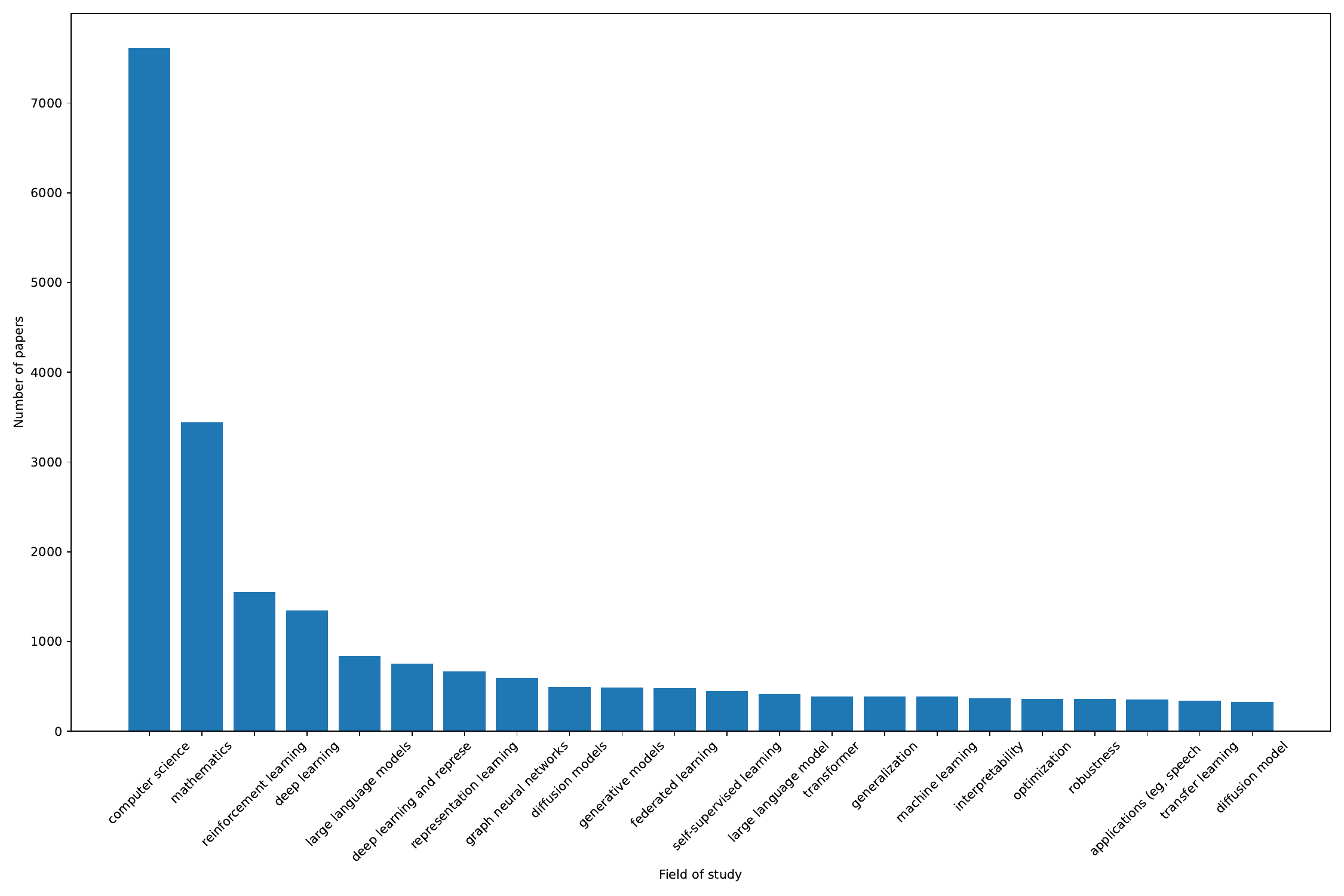}
    \caption{Frequency of the 20 most frequent field of studies in OpenReview.}
    \label{fig:keyword_distribution}
\end{figure}

Citation prediction has been shown to benefit from classifiers trained for specific research fields \cite{citation_prediction_sciencometrics} instead of a single classifier for all data. While the ACL-OCL dataset is already restricted to the field of Computational Linguistics, the OpenReview dataset covers a broader set of topics. Submissions often include a \textit{Field of Study} attribute. However, the \textit{Field of Study} attribute varies widely, containing broad categories like "Computer Science" to more narrow research fields (see Table \ref{tab:research_fields_iclr_2023}). An overview of the distribution of values for the field is presented in Figure \ref{fig:keyword_distribution}. Since the ACL-OCL dataset already focuses on a narrow domain ("Computational Linguistics") we do not perform additional topic classification. To test whether topic labels can improve comparison accuracy of pairwise score prediction models, we label all ICLR submissions with a topic label.

Various classification systems for scientific papers exist \cite{citation_prediction_sciencometrics}, and different models have been explored \cite{benchmark_scientific_document_classification}. Scholarly databases like Semantic Scholar and ArXiv often provide coarse-grained classifications. For example, almost all OpenReview submissions are labeled as \textit{Computer Science} by Semantic Scholar, while ArXiv categorizes them as \textit{cs.AI (Artificial Intelligence)}. Other systems, such as Thomson Reuters’ Web of Science and CSRankings\footnote{\url{https://csrankings.org/index?all&us}}, face similar limitations. Therefore, we adopt an unsupervised classification approach using Latent Dirichlet Allocation (LDA) \cite{lda_orig}. 

We combine titles and abstracts as a unique string, and we pre-process the obtained texts using a common pipeline employing the gensim \cite{gensim} and nltk \cite{nltk} libraries: we tokenize the texts and lemmatize the results. In addition, we also generate bigrams and trigrams, as they are widely used in scientific style. For the number of topics we choose the same number as in ICLR-2023 which is 13. 

\begin{table*}[]
    \centering
    \begin{tabular}{l | l }
         \# Samples & Words \\
         \hline
         6244 & datum, distribution, model, sample, training, generalization, use,  show, method, 
         prediction \\
         5934 & adversarial, model, attack, training, robustness, learning, datum, robust, privacy, client \\
         2753 & network, neural, function, deep, show, gradient, linear, learn, use, matrix  \\ 
         2271  & graph, model, transformer, attention, task, performance, training, node, propose, layer \\
         1888 &  causal, graph, variable, fairness, algorithm, game, tree, structure, effect, decision \\
    \end{tabular}
    \caption{The top 10 most important words for the five most frequent topic resulting from the LDA performed on ICLR submissions.}
    \label{tab:my_label}
\end{table*}

\begin{table*}[]
    \centering
    \begin{tabular}{l | l}
      Venue   & Field of Study  \\
      \hline
      \makecell[l]{ICLR \\2023}  &  \makecell[l]{Unsupervised and Self-Supervised Learning \\
                                Theory (eg. control theory, learning theory, algorithmic game theory) \\
                                \makecell[l]{Social Aspects of Machine Learning \\(eg. AI safety, fairness, privacy, interpretability, human-AI interaction, ethics)} \\
                                Reinforcement Learning (eg. robotics, planning, hierarchical RL,decision and control) \\
                                Probabilistic Methods (eg. variational inference, causal inference, Gaussian processes) \\
                                Optimization (eg. convex and non-convex optimization) \\
                                Neuroscience and Cognitive Science (e.g., neural coding, brain-computer interfaces) \\
                                \makecell[l]{Machine Learning for Sciences \\ (eg. biology, physics, health sciences, social sciences, climate sustainability)} \\
                                Infrastructure (eg. datasets, competitions, implementations, libraries) \\
                                Generative models \\
                                General Machine Learning (ie. none of the above) \\
                                Deep Learning and representational learning \\
                                Applications (eg. speech processing, computer vision, NLP)
                                }
    \end{tabular}
    \caption{Research fields available for authors to self-select when categorizing their submissions for ICLR 2023.}
    \label{tab:research_fields_iclr_2023}
\end{table*}

\section{Qualitative Analysis}\label{app:qualitative_analysis}

\begin{figure*}
    \centering
    \includegraphics[width=\linewidth]{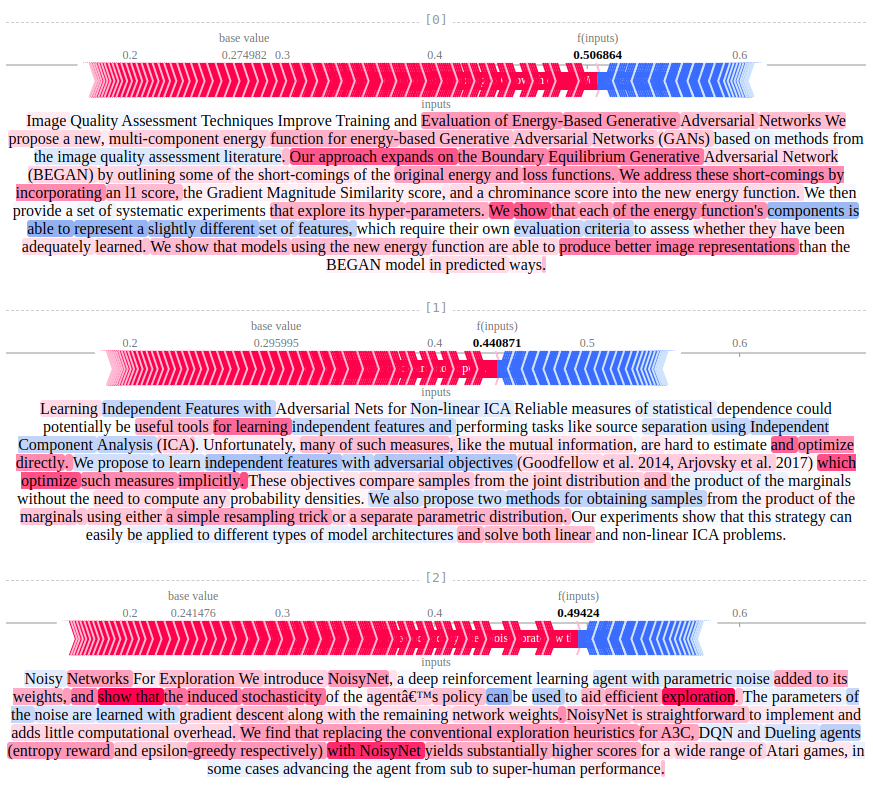}
    \caption{Illustrative Shapley values for titles and abstracts in the review score prediction model trained on the ICLR subset of OpenReview.}
    \label{fig:title_abstract_review_score}
\end{figure*}

\begin{figure*}
    \centering
    \includegraphics[width=\linewidth]{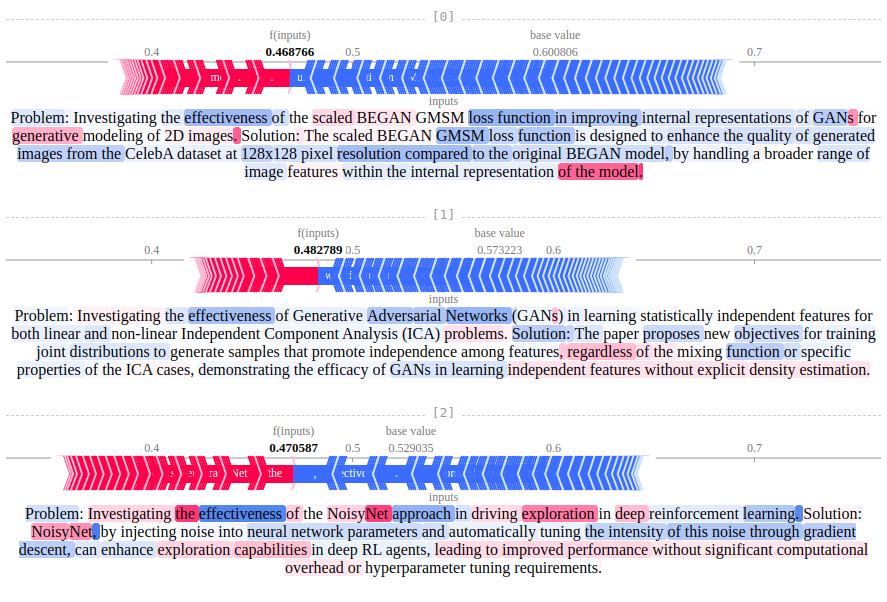}
    \caption{Illustrative Shapley values for research hypotheses in the review score prediction model trained on the ICLR subset of OpenReview.}
    \label{fig:hypothesis_review_score}
\end{figure*}

\begin{figure*}
    \centering
    \includegraphics[width=\linewidth]{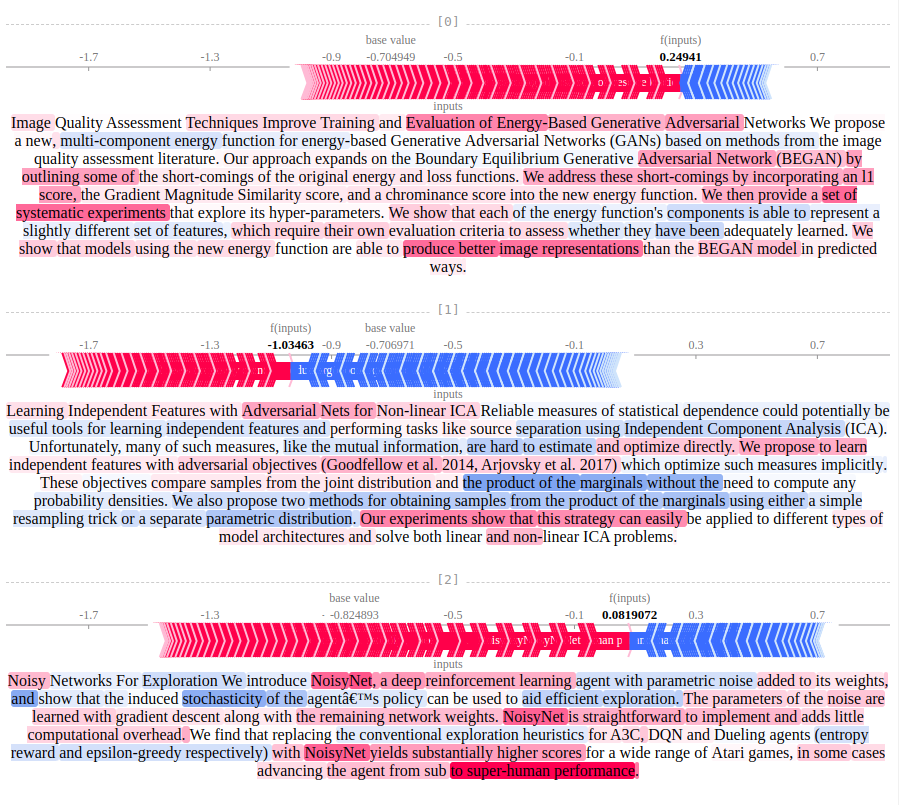}
    \caption{Illustrative Shapley values for titles and abstracts in the citation score prediction model trained on the ICLR subset of OpenReview.}
    \label{fig:title_abstract_citation_score}
\end{figure*}

\begin{figure*}
    \centering
    \includegraphics[width=\linewidth]{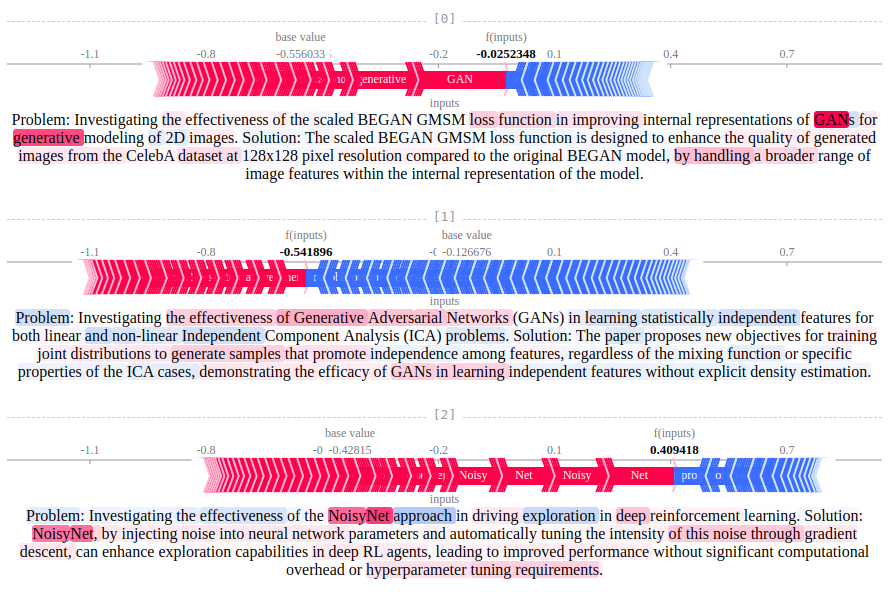}
    \caption{Illustrative Shapley values for research hypotheses in the citation score prediction model trained on the ICLR subset of OpenReview.}
    \label{fig:hypothesis_citation_score}
\end{figure*}

For the qualitative analysis we approximate Shapley values by determining Owen values \cite{owen_values} via Partition Shap\footnote{\url{https://shap.readthedocs.io/en/latest/generated/shap.PartitionExplainer.html}} on the training set. We visualise the approximate Shapley values for three examples for the review score prediction models using title and abstracts (see Figure \ref{fig:title_abstract_review_score}) and hypotheses (see Figure \ref{fig:hypothesis_review_score}) as well as the citation prediction models using title and abstracts (see Figure \ref{fig:title_abstract_citation_score}) and hypotheses (see Figure \ref{fig:hypothesis_citation_score}).

\end{document}